\documentclass{ieeeaccess}

\usepackage{cite}
\usepackage{amsmath,amssymb,amsfonts}
\usepackage{pifont}

\usepackage{bm}
\usepackage{graphicx}
\usepackage{textcomp}
\usepackage{booktabs}
\usepackage{multirow}
\usepackage{array}

\usepackage{enumitem}
\setlist{nolistsep}
\usepackage{adjustbox}
\usepackage{mathtools}
\usepackage{amssymb}
\usepackage{algorithm}
\usepackage{algpseudocode}
\def\x{{\mathbf x}}

\DeclarePairedDelimiter\floor{\lfloor}{\rfloor}
\def\BibTeX{{\rm B\kern-.05em{\sc i\kern-.025em b}\kern-.08em
    T\kern-.1667em\lower.7ex\hbox{E}\kern-.125emX}}
\begin{document}
\history{Date of publication xxxx 00, 0000, date of current version xxxx 00, 0000.}
\doi{10.1109/ACCESS.2017.DOI}

\title{A Unified Deep Learning Framework for Short-Duration Speaker Verification in Adverse Environments}

\author{\uppercase{Youngmoon Jung}, 
\uppercase{Yeunju Choi, Hyungjun Lim, and Hoirin Kim},
\IEEEmembership{Member, IEEE}}
\address{School of Electrical Engineering, Korea Advanced Institute of Science and Technology, Daejeon 34141, South Korea}
\tfootnote{This research was supported by the Korean MSIT (Ministry of Science and ICT), under the National Program for Excellence in SW (2016-0-00018), supervised by the IITP (Institute for Information \& communications Technology Planning \& Evaluation).}

\markboth
{Author \headeretal: Preparation of Papers for IEEE TRANSACTIONS and JOURNALS}
{Author \headeretal: Preparation of Papers for IEEE TRANSACTIONS and JOURNALS}

\corresp{Corresponding author: Youngmoon Jung (dudans@kaist.ac.kr).}

\begin{abstract}
Speaker verification (SV) has recently attracted considerable research interest due to the growing popularity of virtual assistants.
At the same time, there is an increasing requirement for an SV system: it should be robust to short speech segments, especially in noisy and reverberant environments.
In this paper, we consider one more important requirement for practical applications: the system should be robust to an audio stream containing long non-speech segments, where a voice activity detection (VAD) is not applied. 
To meet these two requirements, we introduce feature pyramid module (FPM)-based multi-scale aggregation (MSA) and self-adaptive soft VAD (SAS-VAD).
We present the FPM-based MSA to deal with short speech segments in noisy and reverberant environments.
Also, we use the SAS-VAD to increase the robustness to long non-speech segments.
To further improve the robustness to acoustic distortions (i.e., noise and reverberation), we apply a masking-based speech enhancement (SE) method.
We combine SV, VAD, and SE models in a unified deep learning framework and jointly train the entire network in an end-to-end manner.
To the best of our knowledge, this is the first work combining these three models in a deep learning framework. 
We conduct experiments on Korean indoor (KID) and VoxCeleb datasets, which are corrupted by noise and reverberation. 
The results show that the proposed method is effective for SV in the challenging conditions and performs better than the baseline \textit{i}-vector and deep speaker embedding systems.

\end{abstract}


\titlepgskip=-15pt

\maketitle

\section{Introduction}
\label{sec:introduction}
\IEEEPARstart{S}{peaker} verification (SV) is the task of verifying that an input utterance is spoken by a claimed speaker. 
SV can be classified into two categories: text-dependent SV (TD-SV) and text-independent SV (TI-SV). In TD-SV, the speech content should be the same in the enrollment and verification utterances, while in TI-SV, there are no constraints on the contents of the utterances \cite{Hansen2015}.
Even though TI-SV is more challenging than TD-SV because of the phonetic variability, TI-SV is more convenient from a user point of view in that the user can speak freely to the system.

Over the past decades, the \textit{i}-vector approach \cite{Dehak2011} with probabilistic linear discriminant analysis (PLDA) \cite{Ioffe2006} has been widely used for TI-SV \cite{Kenny2010, Garcia2011, Kenny2013, Zhang2019}. The \textit{i}-vector approach learns a low-dimensional representation containing both speaker and channel variability, through which a variable-length utterance can be represented as a fixed-dimensional \textit{i}-vector. PLDA techniques are used to compensate for the speaker and channel variability of \textit{i}-vectors. 
The \textit{i}-vector/PLDA systems perform well on long enrollment/test utterances (usually more than 10 s), but are prone to have performance degradation on short enrollment/test utterances (usually less than 10 s) \cite{Kanagasundaram2011}.

With the development of deep learning, a deep neural network (DNN)-based acoustic model has been integrated into the \textit{i}-vector/PLDA system and used to generate senone posteriors for \textit{i}-vector computation instead of the conventional Gaussian Mixture Model-Universal Background Model (GMM-UBM) \cite{Kenny2014, Lei2014}.
This approach, called DNN/\textit{i}-vector, improves the GMM-UBM-based \textit{i}-vector system. However, it requires well-annotated training data, and the introduction of the additional DNN-based acoustic model significantly increases the computational complexity.

Another deep-learning-based approach is deep speaker embedding learning, which is the most extensively studied approach. It learns speaker embeddings using speaker features extracted from a speaker-discriminative network \cite{Variani2014, Heigold2016, Zhang2018, Snyder2018}. 
Several convolutional neural networks (CNNs) such as time-delay neural network (TDNN) \cite{Peddinti2015}, 
VGG \cite{Simonyan2015}, and ResNet \cite{He2016} are mostly used in this approach.
Typically, the network is trained to classify speakers in the training set \cite{Variani2014, Snyder2018} or to maximize the distance between same-speaker and different-speaker utterance pairs \cite{Heigold2016, Zhang2018}.
Then, we obtain an utterance-level speaker embedding, named deep speaker embedding, by aggregating speaker features from the network.
The \textit{d}-vector \cite{Variani2014} and \textit{x}-vector \cite{Snyder2018} methods are examples of this approach.

Recently, SV has gained a lot of research interest with the advancement and popularity of virtual assistants such as Amazon Alexa, Apple Siri, Google Assistant, and Microsoft Cortana. 
From application point of view, there is an increasing requirement for SV systems: the systems should be robust to short speech segments.
Otherwise, the user will be asked to speak for a long time during the enrollment and verification phases, thereby causing inconvenience to the user.
To improve performance on short speech segments, several techniques have been proposed in previous studies \cite{Li2018, Huang2018, Wang2019, Hajavi2019, JungJW2019, Jung2020, Kye2020}. 
At the same time, the systems are expected to be robust to noisy and reverberant environments, where they are typically used.
Recent studies have suggested speech enhancement (SE) algorithms to improve the robustness of the SV systems to noise and reverberation \cite{Al-Ali2017, Shon2019, Nidadavolu2020, Kataria2020, JungMH2020}.

In this study, we consider one more requirement which has not been considered in recent SV studies, despite its importance in real-world applications: the SV systems should be robust to the input audio containing long non-speech segments, especially in noisy and reverberant environments.
This assumes that voice activity detection (VAD) has not been applied to remove non-speech frames, which may degrade the SV performance, from the audio.
Most SV studies still rely on a traditional energy-based VAD \cite{Zhang2018, Cai2018, Wang2019}, and even some of them do not apply VAD \cite{Nagrani2017, Jung2019}. 
It is because most SV databases are included in the following two cases, thus minimizing the need of robust VAD: (1) They were recorded in relatively clean conditions, where the naive energy-based VAD performs reasonably well. (2) They contain audio recordings which already have small portion of non-speech.
However, our previous work \cite{Jung2019VAD} shows the need of the robust VAD for SV in real-world environments, where the input audio contains long non-speech segments in noisy and reverberant environments.
In these adverse environments, the energy-based VAD produces unreliable speech frames, which degrades the performance of SV systems \cite{Mandasari2012}. 

To satisfy these two requirements for TI-SV, which is our ultimate goal in this paper, we present our methods: feature pyramid module (FPM)-based multi-scale aggregation (MSA) \cite{Jung2020} and self-adaptive soft VAD (SAS-VAD) \cite{Jung2019VAD}.
We employ the FPM-based MSA to deal with short speech segments.
Also, we adopt the SAS-VAD to deal with long non-speech segments.
In the experiments, we show that both algorithms are robust to acoustic distortions (i.e., noise and reverberation).
We further improve the SAS-VAD and combine it with the FPM-based MSA.
Finally, we integrate a masking-based SE model into the combined model, thus further increasing the robustness to the acoustic distortions.
We jointly train the entire network in an end-to-end manner. 
Our end-to-end approach has advantages over the conventional approach using separately pre-trained models. 
As the VAD and SE models are optimized to minimize the SV loss, they do not require labels for training, which are difficult to obtain for most SV datasets. Moreover, they are guided by the SV loss to generate outputs which are more suitable and useful for the SV task.  
To the best of our knowledge, this is the first work that combines SV, VAD, and SE models into a unified deep learning framework.

The main contributions of this paper are:
(1) We provide a comprehensive overview of deep speaker embedding learning, including its loss functions, operation types, and pooling methods. 
(2) We present a new practical consideration for SV systems that has never been discussed before: short-duration SV with long non-speech segments in noisy and reverberant environments.
(3) To achieve our goal, we combine the three approaches: FPM-based MSA, SAS-VAD, and masking-based SE, and the whole network is trained in an end-to-end manner. 
Especially, we propose a 1D-CNN-based synchronizer to combine FPM-based MSA with SAS-VAD. Besides, we conduct extensive experiments using different types of feature extractors and acoustic features.

The remainder of this paper is organized as follows. 
An overview of deep speaker embedding learning is presented in Section \ref{sec:deep_speaker_embedding}.
FPM-based MSA and masking-based SE are introduced in Section \ref{sec:multi-scale_aggregation} and Section \ref{sec:speech-enhancement}, respectively. 
SAS-VAD algorithm and
proposed combined model are presented in Section \ref{sec:SAS-VAD} and Section \ref{sec:integrated_method}, respectively.
The experimental setup is described in Section \ref{sec:DB_systems}.
The experimental results on different datasets are given in Section \ref{sec:results}.
Finally, we summarize our work and draw conclusions in Section \ref{sec:conclusion}.

\section{Deep speaker embedding learning}
\label{sec:deep_speaker_embedding}
In this section, we provide an extensive overview of the deep speaker embedding learning. 
As mentioned in Section \ref{sec:introduction}, deep-learning-based SV approaches can be divided into two types: DNN/\textit{i}-vector and deep speaker embedding learning. 
Wang \textit{et al.} \cite{Wang2019} denote the former as cascade embedding learning and the latter as direct embedding learning. 
In this work, we focus on the deep speaker embedding learning due to the limitations of DNN/\textit{i}-vector and the increasing popularity of deep speaker embedding learning.

We can categorize deep speaker embedding learning according to the loss function. 
The first approach is based on the softmax loss, which is defined in \cite{Liu2016} as the combination of a cross-entropy loss, a softmax function, and the last fully-connected layer \cite{Variani2014, Snyder2018}. 
In this approach, a network is trained to classify speakers in the training set. 
The second one is based on the metric learning based loss, such as triplet loss. 
This loss encourages the intra-class compactness and inter-class separability, thereby improving generalization performance \cite{Heigold2016, Zhang2018}. 
A disadvantage is that we have to select triplets from the training set carefully, which is both performance-sensitive and time-consuming.
To overcome this problem and learn more discriminative embeddings, advanced classification-based losses, such as center loss \cite{Wen2016} and angular softmax (A-Softmax) loss \cite{Liu2017}, are applied to SV \cite{Li2018, Huang2018}. 
The center loss minimizes the Euclidean distance between embeddings and their corresponding class centers. 
The A-Softmax loss introduces an angular margin into the softmax loss, enhancing the discriminability of embeddings.
More advanced angular margin losses have been proposed, such as additive margin softmax \cite{Wang2018} and additive angular margin softmax \cite{Deng2019} losses, achieving state-of-the-art performance on the SV task \cite{Liu2019}.

\Figure[!t]
(topskip=0pt, botskip=0pt, midskip=0pt)[trim=1.5cm 0.2cm 1.5cm 0.1cm, clip=true, width=0.45\textwidth]{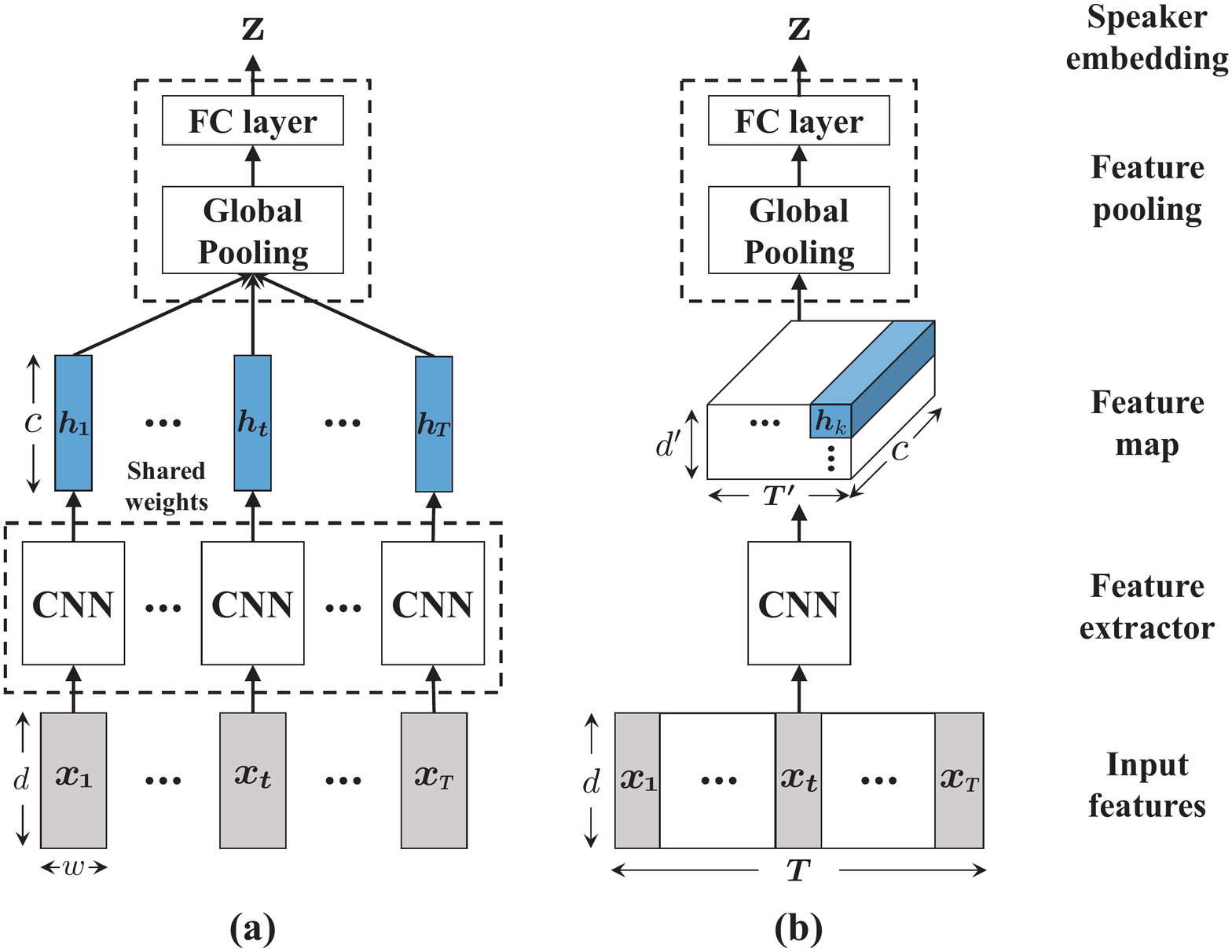}
   {\textbf{Two types of operations in deep speaker embedding learning: (a) frame-level operation and (b) segment-level operation.}\label{fig:two_operations}} 
   
In training, a fixed-length segment is randomly selected from the input utterance and the segment is fed into the network. 
The first reason is that the length of the input utterance can be extremely long with limited GPU memory. The second one is that, to form a mini-batch, the size of all samples in the mini-batch must be the same.
After training, we feed an entire utterance into the network to extract a speaker embedding and the embedding is stored for each enrollment speaker.
Finally, scoring between enrollment and test embeddings is performed using either the cosine similarity or PLDA. Based on how the network operates on input speech segments, the deep speaker embedding learning can be classified into two types: frame-level and segment-level operations, which are illustrated in Fig. \ref{fig:two_operations}(a) and (b), respectively.

In frame-level operation, each acoustic feature vector $\bm{x}_t\in\mathbb{R}^{d}$ in the input segment is augmented with neighboring frames within a context window of size $w$ and fed into a CNN-based feature extractor.
Here, $t$ is the frame index.
Then, a frame-level speaker feature vector $\bm{h}_t\in{\mathbb{R}}^{c}$ is extracted from the feature extractor for each frame to form a 2D feature map $\bm{H} = [\bm{h}_1 \, \bm{h}_2 \, \cdots \, \bm{h}_T] \in{\mathbb{R}}^{c\times T}$. 
Here, $T$ is the total number of frames in the input segment.
Note that the number of $\bm{x}_t$ and $\bm{h}_t$ are the same.
After extracting the feature map, we apply feature pooling to map the variable-length feature map $\bm{H}$ to a fixed-dimensional speaker embedding $\bm{z}$.
For feature pooling, we first aggregate the feature vectors across time by using global pooling
and then obtain a pooled feature vector. The pooled vector is passed to one or few fully-connected (FC) layers to generate the deep speaker embedding $\bm{z}$. 
The works in \cite{Snyder2018, Huang2018,Wang2019,Jung2019VAD} are examples of this approach.

In segment-level operation, all $\bm{x}_t\in\mathbb{R}^{d}\,\,(t=1,...,T)$ in each segment are combined to form a feature matrix $\bm{X}\in{\mathbb{R}}^{d\times T}$ and then fed into the feature extractor at once. We denote the resulting 3D feature map as $\bm{H}\in{\mathbb{R}}^{d^{\prime} \times T^{\prime} \times c}$, where $c$ is the channel dimension of the last convolutional layer.
Note that $d^{\prime}$ and $T^{\prime}$ are smaller than $d$ and $T$, respectively, due to the repeated local pooling operations.
Here, there are two ways to aggregate speaker feature vectors $\bm{h}_k\in{\mathbb{R}}^{c}\,\,(k=1,...,d^{\prime}T^{\prime})$ into a single feature vector. The first one is to aggregate feature vectors across both time and frequency.
The studies in \cite{Bhattacharya2017, Li2018, Jung2020, Jung2019} are examples of this approach. The second one is to reduce the frequency dimension $d^{\prime}$ to 1 by additional global pooling \cite{Cai2018} or FC layers \cite{Nagrani2017} before applying the global pooling. 
After reducing the dimension, the global pooling aggregates feature vectors across time, which is the same as in the frame-level operation. In this work, we use the former approach.

Meanwhile, we can divide pooling operations into two types in terms of receptive field: local pooling and global pooling \cite{Kobayashi2019}. 
In local pooling, the pooling block is smaller than the input feature map and the time-frequency scale is reduced with increasing robustness against temporal and spectral variations in input speech.
Different from local pooling, global pooling covers the entire input feature map and compresses the feature map into a feature vector of size $c$.
Therefore, in deep speaker embedding learning, local pooling is commonly used in a feature extractor to extract useful speaker features, and global pooling is used in a feature pooling layer to aggregate speaker feature vectors into a pooled feature vector. 

The global average pooling (GAP) is the most naive method for global pooling \cite{Snyder2016, Nagrani2017}.
Recently, many researchers have proposed advanced pooling methods for deep speaker embedding learning.
Snyder \textit{et al.} \cite{Snyder2017} introduce the statistics pooling (SP) where the standard deviation of the feature vectors is used as well as the average of them. 
Okabe \textit{et al.} \cite{Okabe2018} present the attentive statistics pooling (ASP), which integrates attention mechanism into the statistics pooling.
Zhang \textit{et al.} \cite{Zhang2018} propose to use the spatial pyramid pooling (SPP) \cite{He2014}, which divides the last feature map into several bins and applies GAP to each bin.
Cai \textit{et al.} \cite{Cai2018} apply the learnable dictionary encoding (LDE), which imitates the process of encoding GMM supervectors \cite{Campbell2006} within a deep learning framework. 
Jung \textit{et al.} \cite{Jung2019} propose the spatial pyramid encoding (SPE), which improves both LDE and SPP methods by combining them.

For all the pooling methods mentioned above, we use only single-scale feature map from the last layer of the feature extractor. 
Recently, multi-scale aggregation (MSA) methods have been proposed to exploit speaker information at multiple time scales  \cite{Gao2019, Hajavi2019, Jung2020, Seo2019}, showing the effectiveness in dealing with variable-duration test utterances.

\Figure[!t]
(topskip=0pt, botskip=0pt, midskip=0pt)[trim=0.0cm 5.5cm 0.0cm 0.95cm, clip=true, width=0.42\textwidth]{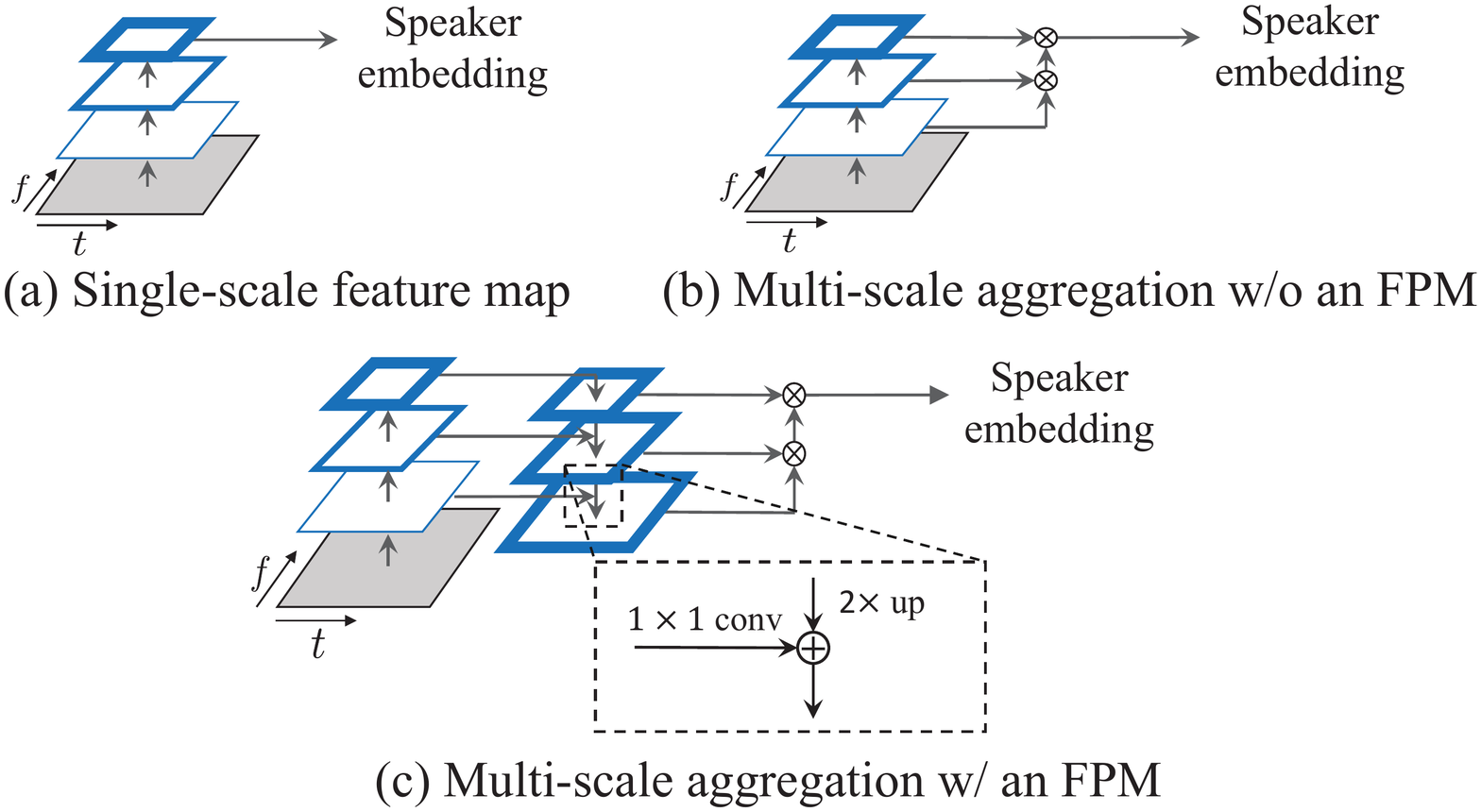}
   {\textbf{Three types of deep speaker embedding learning. (a) Using only single-scale feature maps. (b) Using multi-scale feature maps without a feature pyramid module (FPM). (c) Using multi-scale feature maps with an FPM. In this paper, input feature matrix $\bm{X}$ is represented by grey rectangles and CNN feature maps are marked by blue outlines.
   Thicker outlines correspond to more speaker-discriminative information in feature maps. $\otimes$ : concatenation, $\oplus$ : element-wise addition, 2$\times$ up : upsampling operation by a factor of 2.}\label{fig:PFH_FPN}}

\section{Multi-scale aggregation}
\label{sec:multi-scale_aggregation}
As mentioned in Section \ref{sec:introduction}, deep CNNs are commonly used as a feature extractor for deep speaker embedding. 
Deep CNNs are generally bottom-up and feed-forward architectures, consisting of alternating layers of convolution and local pooling 
to learn discriminative features, which operates at the segment-level (Fig. \ref{fig:two_operations}(b)).
By doing so, deep CNNs compute a feature hierarchy layer by layer, which is inherently multi-scale of pyramidal shape due to repeated local pooling layers.
This in-network feature hierarchy produces feature maps of different time-frequency scales and resolutions, but it also produces large semantic gaps between different layers.
In deep speaker embedding learning, as the feature extractor is trained to discriminate speakers, the features from higher layers contain higher-level speaker information (i.e., more speaker-discriminative) \cite{Wang2017} but have smaller scales (i.e., lower resolutions) than those from lower layers.

Thanks to the local pooling operations, deep CNNs are robust to scale variation, thus making it possible to use feature maps computed on a single input scale (Fig. \ref{fig:PFH_FPN}(a)).
Even with this robustness, using multi-scale features from multiple layers (Fig. \ref{fig:PFH_FPN}(b)), called multi-scale aggregation (MSA), has shown better performance than using single-scale feature maps \cite{Gao2019, Hajavi2019, Jung2020, Seo2019}. 
Note that, between the frame- and segment-level operations, we should choose the segment-level operation for the MSA because all the feature maps from different layers have the same time scale in the frame-level operation.
To improve the MSA, we propose to use a feature pyramid module (FPM) which is illustrated in Fig. \ref{fig:PFH_FPN}(c). 
In the following, we review related works and discuss the relation between our approach and previous ones.

\Figure[!t]
(topskip=0pt, botskip=0pt, midskip=0pt)[trim=0.1cm 2.8cm 0.1cm 0.4cm, clip=true, width=0.4\textwidth]{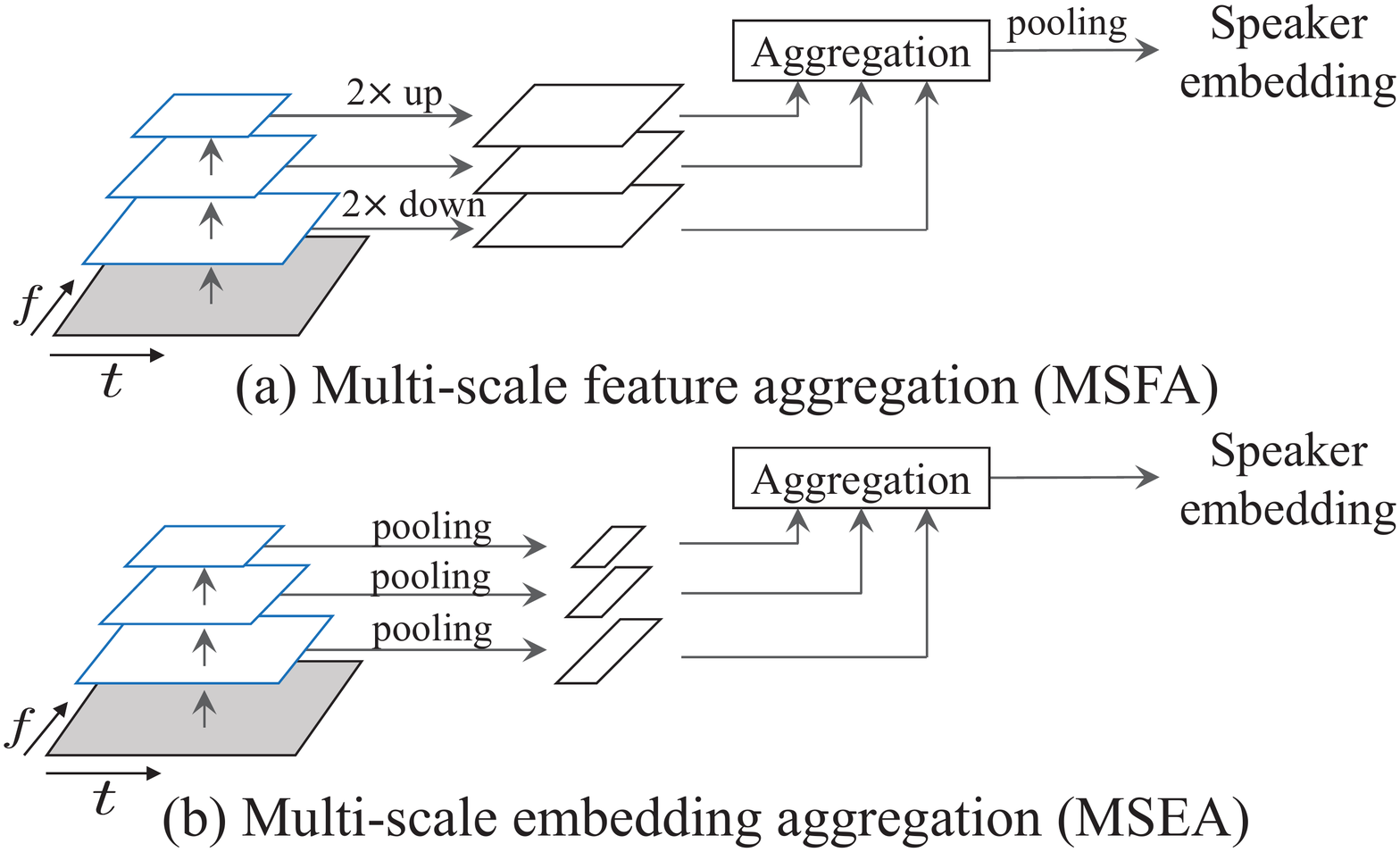}
   {\textbf{Two types of multi-scale aggregation (MSA). (a) Multi-scale feature aggregation (MSFA). (b) Multi-scale embedding aggregation (MSEA). ``2$\times$ down'' denotes the downsampling operation by a factor of 2. In all figures, ``pooling'' denotes global pooling operation.}\label{fig:MSFA_MSEA}}

\subsection{Related works}

Gao \textit{et al.} \cite{Gao2019} proposed multi-stage aggregation for deep speaker embedding learning. 
They used ResNet as a feature extractor, where the feature maps of stage 2, 3, and 4 (see Table \ref{SR_architecture}) were concatenated along the channel axis. 
Before concatenation, they downsampled and upsampled the feature maps of different sizes (i.e., different time-frequency scales) to make them have the same size.
Concretely, the feature map of stage 2 was downsampled by convolution with stride 2, and the feature map of stage 4 was upsampled by bilinear interpolation or transposed convolution. 
After concatenation, statistics pooling was applied to generate speaker embeddings. 
In this approach, speaker embeddings were obtained using feature maps at multiple time-frequency scales, achieving state-of-the-art performance on VoxCeleb \cite{Nagrani2017}.

Seo \textit{et al.} \cite{Seo2019} also utilized features from different stages of ResNet to combine information at different time-frequency scales.
Different from the approach of Gao \textit{et al.}, GAP was applied to the feature maps respectively and the resulting pooled feature vectors were concatenated into a long vector. The concatenated vector was fed into fully-connected layers to generate the speaker embedding. 
Hajavi \textit{et al.} \cite{Hajavi2019} proposed a similar approach using UtterIdNet to deal with short speech segments.
They showed that the MSA is useful for short-duration speaker verification by extracting as much information as possible from short speech segments.
   
For convenience, we denote the first approach as multi-scale feature aggregation (MSFA) and the second one as multi-scale embedding aggregation (MSEA), which are shown in Fig. \ref{fig:MSFA_MSEA}.
In \cite{Jung2020}, we showed that the MSEA performs slightly better than the MSFA with fewer parameters. 
Furthermore, the MSEA can flexibly use various number of feature maps from different stages, while the MSFA does not.
Therefore, we only consider the MSEA in this paper. 

\subsection{Feature pyramid module}
\label{sec:FPM}
As explained above, in the deep CNN-based feature extractor, 
feature maps of lower layers contain less speaker-discriminative information than those of higher layers. Intuitively, if we can enhance speaker discriminability of the lower-layer feature maps, the performance of the MSA will improve accordingly. 
Motivated by this, we propose to use an FPM to extract multi-scale feature maps containing sufficient speaker-discriminative information at all layers. 
In this paper, we use the FPM-based MSA to deal with short speech segments, especially in noisy and reverberation environments. Besides, we conduct extensive experiments using different types of networks, acoustic features, and datasets.

\Figure[!t]
(topskip=0pt, botskip=0pt, midskip=0pt)[trim=2.7cm 0.0cm 2.7cm 0.0cm, clip=true, width=0.48\textwidth]{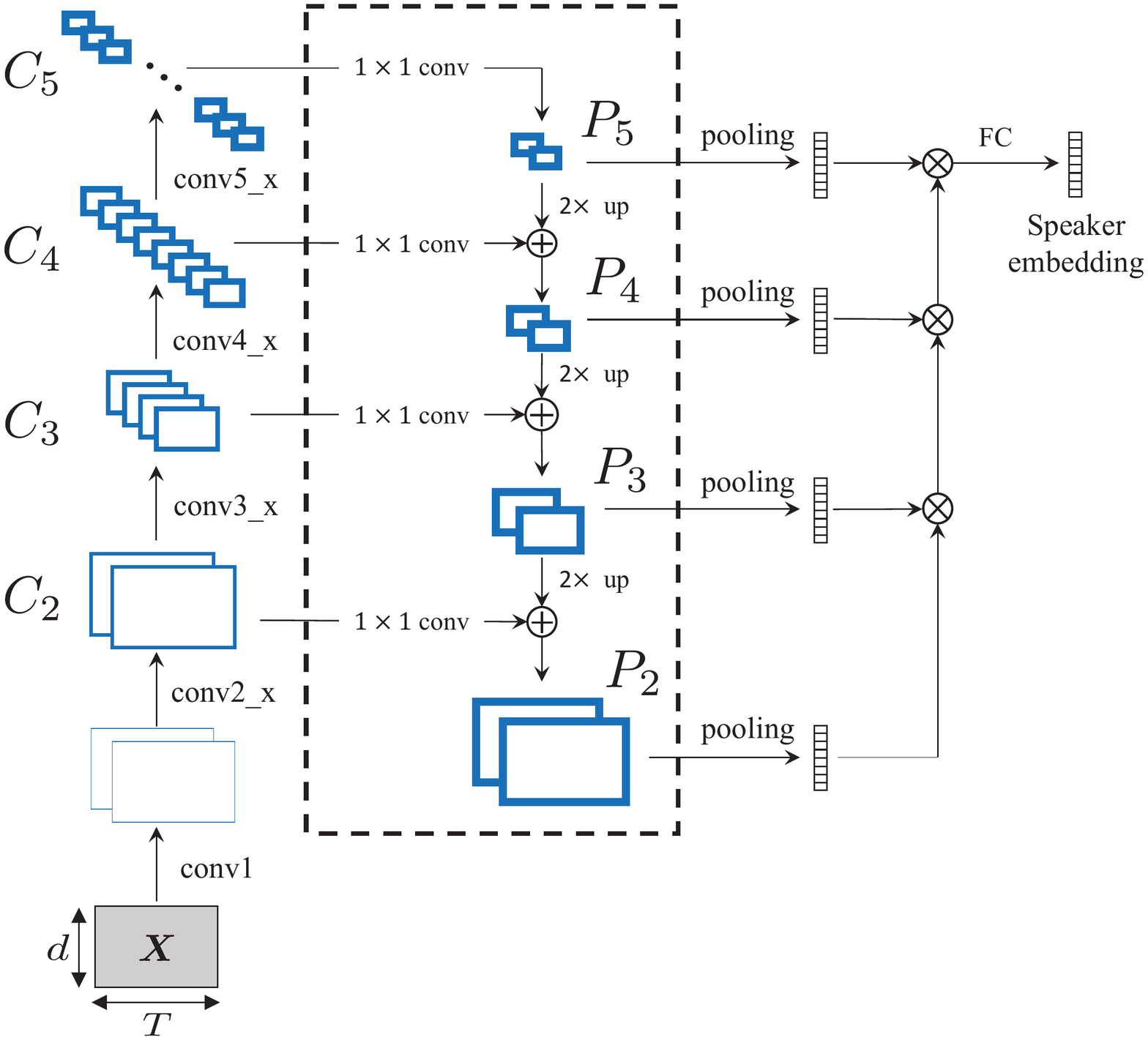}
   {\textbf{Illustration of multi-scale aggregation (MSA) using a feature pyramid module (FPM). The black dotted box indicates the FPM.}\label{fig:PFH_FPN_details}}

The detailed architecture is presented in Fig. \ref{fig:PFH_FPN_details}, which consists of three main components: a bottom-up pathway, a top-down pathway, and lateral connections. 
The bottom-up pathway is a typical feed-forward computation of the feature extractor, which produces feature maps of different scales.
In each ResNet stage, there are several layers generating feature maps of the same time-frequency scale (see Table \ref{SR_architecture}). Since the deepest layer is expected to learn the strongest features, we only choose the output features of the last layer as the output of each stage.
We denote the output of stage $i$ as $C_{i+1}$ for $i=1,\,2,\,3,\,4$, since the stage 1 corresponds to conv2.

The black dotted box in Fig. \ref{fig:PFH_FPN_details} shows the FPM which includes the top-down pathway and the lateral connection. 
The procedure is as follows: (1) At the beginning, a $1\times1$ convolutional layer reduces the channel dimension of $C_5$ to 32 which is the channel dimension of the stage 1. 
(2) In the top-down pathway, we upsample $C_5$ from stage 4 by a factor of 2 by using transposed convolution. 
In other words, the top-down pathway creates a 3D feature map consisting of 2D ``time-frequency (TF)'' feature maps which are larger than those of $C_5$ (note that $C_5$ also consists of several TF feature maps). These TF feature maps have the same size as those of $C_4$, but contain more speaker-discriminative information.
(3) The upsampled feature map is then enhanced by $C_4$ from the bottom-up pathway via lateral connections.
More concretely, the top-down feature map is merged with the corresponding bottom-up feature map by element-wise addition.
Before merging, a $1\times1$ convolution
reduces the channel dimension of $C_4$ to 32.
These lateral connections play the same role as the skip connections in U-Net \cite{Ronneberger2015}. 
(4) The process from step (1) to step (3) is repeated from the top stage to the bottom stage. 
(5) Finally, convolutional layers are added to each merged feature map to reduce the aliasing effect of upsampling. 
Specifically, we first apply a $1\times1$ convolution with 32 filters, and then increase the channel dimension to that of the corresponding bottom-up feature map by using a $3\times3$ convolution.
This final feature map is called $P_{i}$ corresponding to $C_{i}$ for $i=2,\,3,\,4,\,5$, where $P_{i}$ and $C_{i}$ have the same time-frequency resolution.

The FPM enhances higher-resolution feature maps containing lower-level speaker information by providing higher-level speaker information from lower-resolution feature maps.
The resulting feature pyramid has abundant speaker-discriminative information at all stages. 
Furthermore, the FPM reduces the total number of parameters in the network because the channel dimensions of stage 2, 3, and 4 are reduced to 32, which is the minimum number of filters.

According to a recent study \cite{Veit2016}, the collection of variable-length paths through ResNet shows ensemble-like behavior, in that the paths do not heavily depend on each other. Likewise, we can say that multiple paths generated by the MSA use an ensemble of multi-scale features that are extracted from different paths.
As the variable-length feature maps are used to generate speaker embeddings, we expect that the performance of deep speaker embedding learning will be improved for variable-length test utterances. 
In our previous study \cite{Jung2020}, we showed that using the MSA improves the performance for both short and long utterances, and the FPM further enhances the performance of the MSA.

\section{Speech enhancement}
\label{sec:speech-enhancement}

\Figure[!t]
(topskip=0pt, botskip=0pt, midskip=0pt)[trim=0.0cm 5.3cm 0.0cm 5.8cm, clip=true, width=0.48\textwidth]{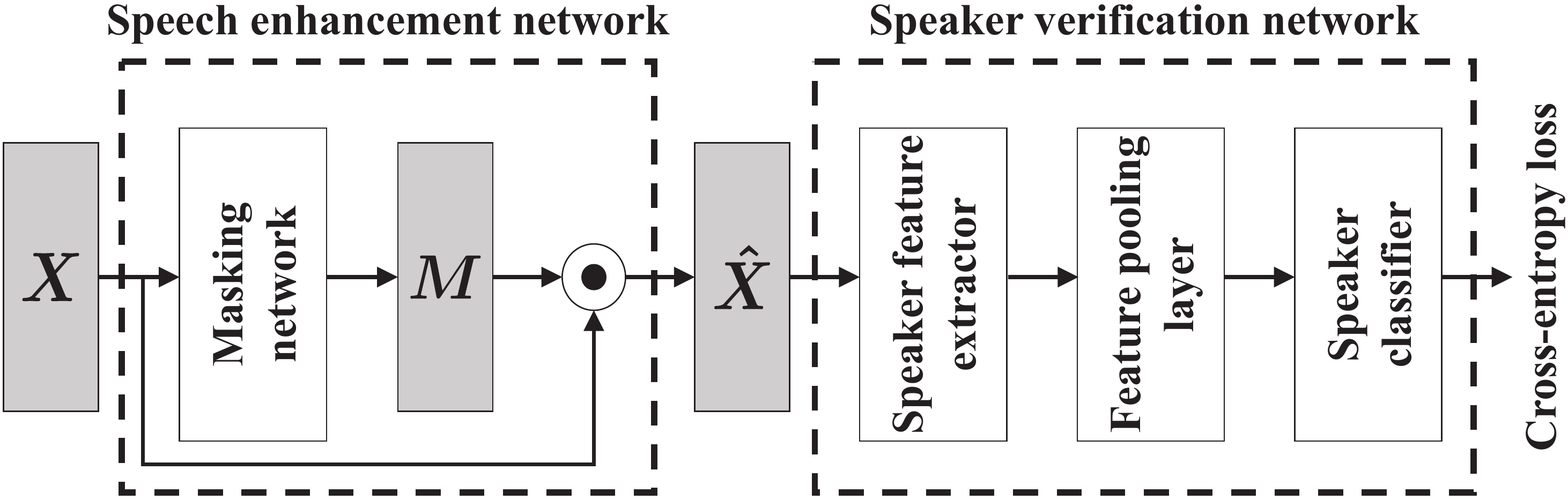}
   {\textbf{Speech enhancement network for robust speaker verification. In this paper, $\odot$ denotes element-wise multiplication.}\label{fig:masking_network}}
  
The FPM-based MSA has the robustness to acoustic distortions, as will be shown in the experiment section. 
To further increase the robustness, we apply speech enhancement using a masking network which consists of 11 dilated convolution layers.
The first 10 layers have 16 filters with kernel size $3\times3$ and dilation size $2\times2$.
After each convolution, batch normalization followed by ReLU is applied.
The last convolutional layer has one filter with kernel size $1\times1$ and dilation size $1\times1$.
To obtain a ratio mask, we use a sigmoid function in the last layer that gives values between 0 and 1. 
Employing dilated convolutions increases the receptive field of the network exponentially, resulting in large temporal context. 
The network estimates the ratio mask $\bm{M}\in{\mathbb{R}}^{d \times T}$ and the resulting mask is multiplied with the corrupted feature matrix $\bm{X}\in{\mathbb{R}}^{d \times T}$ element-wise to produce the enhanced feature matrix $\bm{\hat{X}}=\bm{X} \odot \bm{M} \in{\mathbb{R}}^{d \times T}$. 

In \cite{Shon2019}, the speaker verification network is pre-trained and fixed before the masking network is trained. Instead, in this work, we jointly train the masking and speaker verification networks.
The masking network is trained in an end-to-end manner without an explicit loss function. The detailed structure is shown in Fig. \ref{fig:masking_network}. 
After enhancement, the enhanced feature matrix $\bm{\hat{X}}$ is fed into a speaker verification network. Here, speaker feature vectors are extracted by a feature extractor and converted into a speaker embedding $\bm{z}$ by a feature pooling layer. Finally, the combined network is jointly trained to classify speakers in the training set using cross-entropy loss.
When the SAS-VAD is combined with the FPM-based MSA, $\bm{\hat{X}}$ is fed into both speaker verification and VAD networks to improve the robustness of both networks.
This will be further explained in Section \ref{sec:integrated_method}.

\section{Self-adaptive soft Voice activity detection}
\label{sec:SAS-VAD}
To improve the robustness of the SV model to long non-speech segments, we proposed self-adaptive soft VAD (SAS-VAD) \cite{Jung2019VAD}, which is the combination of soft VAD and self-adaptive VAD. 
Here, we introduce the advanced version of SAS-VAD which shows better performance than the original one and can be combined with the MSA to achieve our ultimate goal.
   
\subsection{Soft VAD}
\label{sec:soft-VAD}
In this subsection, we explain our previous soft VAD \cite{Jung2019VAD} first and its advanced version later.
Unlike typical VADs that make a hard decision on acoustic features with a predefined threshold, the soft VAD makes a soft decision on speaker feature vectors when the self-attentive pooling (SAP) \cite{Cai2018} is applied.
By removing the need to find the optimal threshold to make a speech/non-speech decision, this approach helps to improve the generalization ability of VAD \cite{McLaren2015}. This is because the optimal threshold may differ with test conditions. 

With the soft VAD, each speaker feature vector is weighted by its corresponding speech posterior, which is a confidence measure for speech. 
That is, the soft VAD is applied not to acoustic feature vectors $\bm{x}_t\in{\mathbb{R}}^{d}$, but to speaker feature vectors $\bm{h}_t\in{\mathbb{R}}^{c}$ for $t=1,...,T$.  
Specifically, after extracting a speaker feature map $\bm{H}\in{\mathbb{R}}^{c \times T}$ which consists of $T$ feature vectors $\bm{h}_t$, we multiply each $\bm{h}_t$ by its attention weight $\alpha_t$ and speech posterior $q_t$ together for all $T$ frames. The attention weight is calculated by an attention module and the speech posterior is generated by a VAD network. 
In our previous soft VAD, we use the frame-level operation in Fig. \ref{fig:two_operations}(a), and thus the number of $\bm{h}_t$ is the same as the number of $\bm{x}_t$ as discussed in Section \ref{sec:deep_speaker_embedding}. Therefore, we can also obtain $T$ speech posteriors from the VAD network since the VAD operates in a frame-wise manner. 
Concretely, the VAD network is fed by $\bm{x}_t$ with neighboring frames and outputs the corresponding speech posterior $q_t$.
Obviously, a speech posterior vector $\bm{q}=[q_1,...,q_T]\in{\mathbb{R}}^{T}$ from VAD and a speaker feature map $\bm{H}$ from the feature extractor have the same length $T$, thus we can apply the soft VAD.
However, since all the feature maps from different layers have the same time scale, we cannot apply the MSA in the frame-level operation. 
Hence, to combine soft VAD and MSA, we need to modify our soft VAD framework to enable the segment-level operation in Fig. \ref{fig:two_operations}(b).
When we use the segment-level operation, feature maps from different layers have different time scales. Specifically, as the ResNet stage increases, the length of feature map is halved due to local pooling operations. 

\begin{table}[t]
\centering
\begin{footnotesize}
\renewcommand{\arraystretch}{1}
\caption{\textbf{The architecture of 1D-CNN-based synchronizer.}}
\vspace{-0.2cm}
\label{synchronizer}
\begin{tabular}{c|c|c|c}
\hline
Layer    & Filter size & Stride & \# Channels \\ \hline
conv1\_1 & 3           & 1      & 16       \\ 
conv1\_2 & 3           & 2      & 16       \\ 
conv1\_3 & 1           & 1      & 1        \\ \hline
conv2\_1 & 3           & 1      & 32       \\ 
conv2\_2 & 3           & 2      & 32       \\ 
conv2\_3 & 1           & 1      & 1        \\ \hline
conv3\_1 & 3           & 1      & 64       \\ 
conv3\_2 & 3           & 2      & 64       \\ 
conv3\_3 & 1           & 1      & 1        \\ \hline
\end{tabular}
\vspace{-0.4cm}
\end{footnotesize}
\end{table}

Here, another problem arises: a length mismatch occurs between $\bm{q}$ and $\bm{H}$.
To synchronize the VAD output and reduced speaker feature maps, we propose a 1D-CNN-based synchronizer where the local pooling operations are synchronized with the speaker feature extractor along the time axis.
The detailed architecture is presented in Table \ref{synchronizer}.
The synchronizer consists of 3 convolutional (conv.) blocks and each block consists of 3 conv. layers, where 1D convolution is applied along the time axis.
The first conv. layer increases the number of channels by a factor of 2, and the second conv. layer performs the local pooling operation with stride 2.
Both layers are followed by batch normalization and ReLU activation, respectively.
The final conv. layer reduces the number of channels to 1 with kernel size of 1.
Then, the sigmoid activation function is applied to produce a speech posterior vector which is the reduced version of $\bm{q}$.

To be specific, the synchronizer is fed by the speech posterior vector $\bm{q}\in{\mathbb{R}}^{T}$ and outputs three reduced versions of $\bm{q}$ from conv$\ell\_3$ for $\ell=1,\,2,\,3$. 
We denote the output from conv$\ell\_3$ as $\bm{q}^{(\ell)}$, of which the length is $\floor{\frac{T}{2^{\ell}}}$. 
Here, $\bm{q}^{(0)}$ is equal to $\bm{q}$.
To integrate soft VAD into FPM-based MSA, we expand $\bm{q}^{(i)}$ into a 3D tensor $Q_{i+2}$ with the same size of $P_{i+2}$ for $i=0,\,1,\,2,\,3$ (see Fig. \ref{fig:sync_network}, where the synchronizer is denoted as Sync).
$Q_{i+2}$ is constructed by repeating $\bm{q}^{(i)}$ along the frequency and channel axes.
Then, we multiply $Q_{i+2}$ and $P_{i+2}$ element-wise to perform soft VAD.
By this operation, each feature vector in $P_{i+2}$ is weighted by its corresponding speech posterior which is a confidence measure for speech. 
We denote the resulting feature map as $H_{i+2}$.  
In summary, we obtain speech posterior vectors of reduced length from the synchronizer, which have the same length as their corresponding speaker feature maps. After that, we expand each posterior vector to a 3D tensor and multiply it to its corresponding speaker feature map element-wise. In this way, we can apply soft VAD to multi-scale speaker feature maps obtained by FPM-based MSA.

\Figure[!t]
(topskip=0pt, botskip=0pt, midskip=0pt)[trim=0.0cm 2.5cm 0.0cm 2.6cm, clip=true, width=0.46\textwidth]{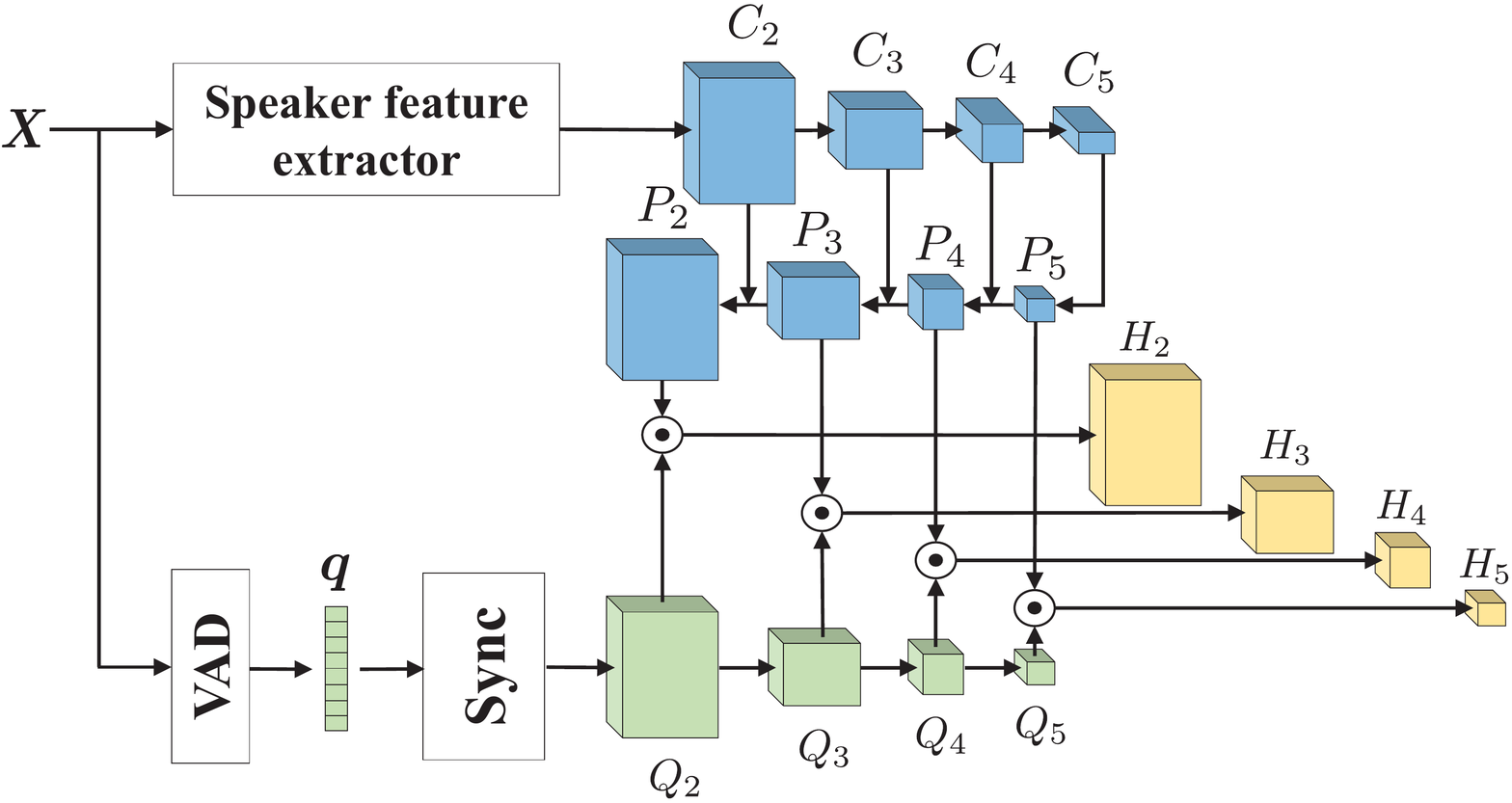}
   {\textbf{The combination of SAS-VAD and FPM-based MSA.}\label{fig:sync_network}}
   
Next, we perform self-attentive pooling (SAP) to obtain a pooled feature vector from the weighted feature map $H_j$ for $j=2,\,3,\,4,\,5$. 
Unlike GAP, where all feature vectors are assumed to be equally important, the SAP layer learns a weight for each feature vector and assigns larger weights to useful feature vectors. The SAP layer itself performs soft VAD implicitly because it gives more weights to feature vectors from speech frames than those from non-speech frames \cite{Zhu2018}. 
Nevertheless, integrating explicit VAD information is found to be helpful for the attention mechanism \cite{Wang2018sofVAD, Jung2019VAD}.

We denote the size of $H_j$ as $d^{(j)} \times T^{(j)} \times c^{(j)}$. $H_j$ corresponds to the stage $j-1$ and consists of feature vectors $\bm{h}_k^{(j)}$ for $k=1,...,d^{(j)}T^{(j)}$, which have the length of $c^{(j)}$.
Henceforth, we omit the superscript $j$ for notational simplicity.
An attention mechanism computes a scalar score $e_{k}$ for $\bm{h}_k$:
\begin{equation}
e_k = {\boldsymbol{v}}^T tanh(\boldsymbol{W}\boldsymbol{h}_{k}+\boldsymbol{b})\,,
\end{equation}
where $\boldsymbol{v}\in{\mathbb{R}}^{c}$ and $\boldsymbol{W}\in{\mathbb{R}}^{c \times c}$ are learnable parameters, and $tanh(\cdot)$ is a tanh activation function. 
Each ResNet stage has its own parameters $\boldsymbol{v}$, $\boldsymbol{W}$, and $\boldsymbol{b}$ in a SAP layer.
Then, we apply a softmax function to normalize the score $e_{k}$:
\begin{equation}
\alpha_k = \frac{exp(e_k)}{\sum_{k'=1}^{dT} exp(e_{k'})}\,.
\end{equation}
After that, the normalized score $\alpha_k$ is used as the weight of $\bm{h}_k$ in global pooling. 
Finally, the weighted mean vector $\boldsymbol{\widetilde{\mu}}$ is calculated as below: 
\begin{equation} \label{eq:att_weightedsum}
\boldsymbol{\widetilde{\mu}} = \sum_{k=1}^{dT} \alpha_k \boldsymbol{h}_k\,.
\end{equation}
In this way, we obtain the pooled vector from each ResNet stage and the following steps are the same as in FPM-based MSA.
All the pooled vectors are concatenated and fed into an FC layer to generate a speaker embedding vector.

\begin{algorithm}
\renewcommand{\algorithmicrequire}{\textbf{Input:}}
\renewcommand{\algorithmicensure}{\textbf{Output:}}
\caption{Self-adaptive soft voice activity detection}
\label{alg:loop}
\begin{algorithmic}[1]
\Require{Training set $D={\{\mathcal{X}^{v}_i,\mathcal{X}^{s}_i,y^{s}_i}\}_{i=1}^U$, pre-trained VAD network $\mathcal{VAD}_0$, posterior threshold $\delta$, loss weight $\lambda$}
\Ensure{Fine-tuned VAD network $\mathcal{VAD}$ with parameters $\theta_v$, speaker verification network $\mathcal{SV}$ with parameters $\theta_s$}
\State $\mathcal{VAD} \gets \mathcal{VAD}_0$
\Repeat{}
    \For{$i := 1$  to  $U$}
    \State // Speech-posterior-based domain adaptation
    \State $\boldsymbol{q}_i \gets \varnothing$
        \For{$t := 1$  to  $T_i$}
            \State{$ q_{i,t} \gets \mathcal{VAD}(\boldsymbol{x}_{i,t}^{v};\theta_{v})$}
            \State $\boldsymbol{q}_i \gets \boldsymbol{q}_i \cup \{q_{i,t}\}$
        \EndFor
        \State $\boldsymbol{\bar{X}}^{v}_i, \boldsymbol{\bar{Y}}^{v}_i \gets \mathcal{F}(\boldsymbol{q}_i, \delta)$
        \State $\mathcal{L}_{SP} \gets \mathcal{L}(\mathcal{VAD}(\boldsymbol{\bar{X}}^{v}_i;\theta_{v}), \boldsymbol{\bar{Y}}^{v}_i)$

    \State // Joint-learning-based domain adaptation
        \State $ \mathcal{L}_{JL} \gets \mathcal{L}(\mathcal{SV}(\boldsymbol{X}^{s}_i;\theta_{s}),y^{s}_i)$
        \State // Calculate losses $\mathcal{L}_{v}$ and $\mathcal{L}_{s}$
        \State $\mathcal{L}_{v} \gets \mathcal{L}_{JL}+\lambda\mathcal{L}_{SP}$
        \State $\mathcal{L}_{s} \gets \mathcal{L}_{JL}$
        \State // Update parameters $\theta_{v}$ and $\theta_{s}$
        \State $\theta_{v} \gets \theta_{v}-\eta_{v} \nabla_{\theta_{v}}\mathcal{L}_{v}$ 
        \State $\theta_{s} \gets \theta_{s}-\eta_{s} \nabla_{\theta_{s}}\mathcal{L}_{s}$ 
    \EndFor
\Until{convergence of $\mathcal{SV}$}

\end{algorithmic}
\label{alg:SSVAD}
\end{algorithm}

In our previous work \cite{Jung2019VAD}, we use a simple fully-connected DNN-based VAD, but in this work, we also use more advanced neural networks for VAD: long short-term memory (LSTM) and convolutional, long short-term memory, fully connected deep neural network (CLDNN) \cite{Zazo2016}. 

\Figure[!t]
(topskip=0pt, botskip=0pt, midskip=0pt)[trim=0.0cm 24.5cm 0.0cm 24.5cm, clip=true, width=0.9\textwidth]{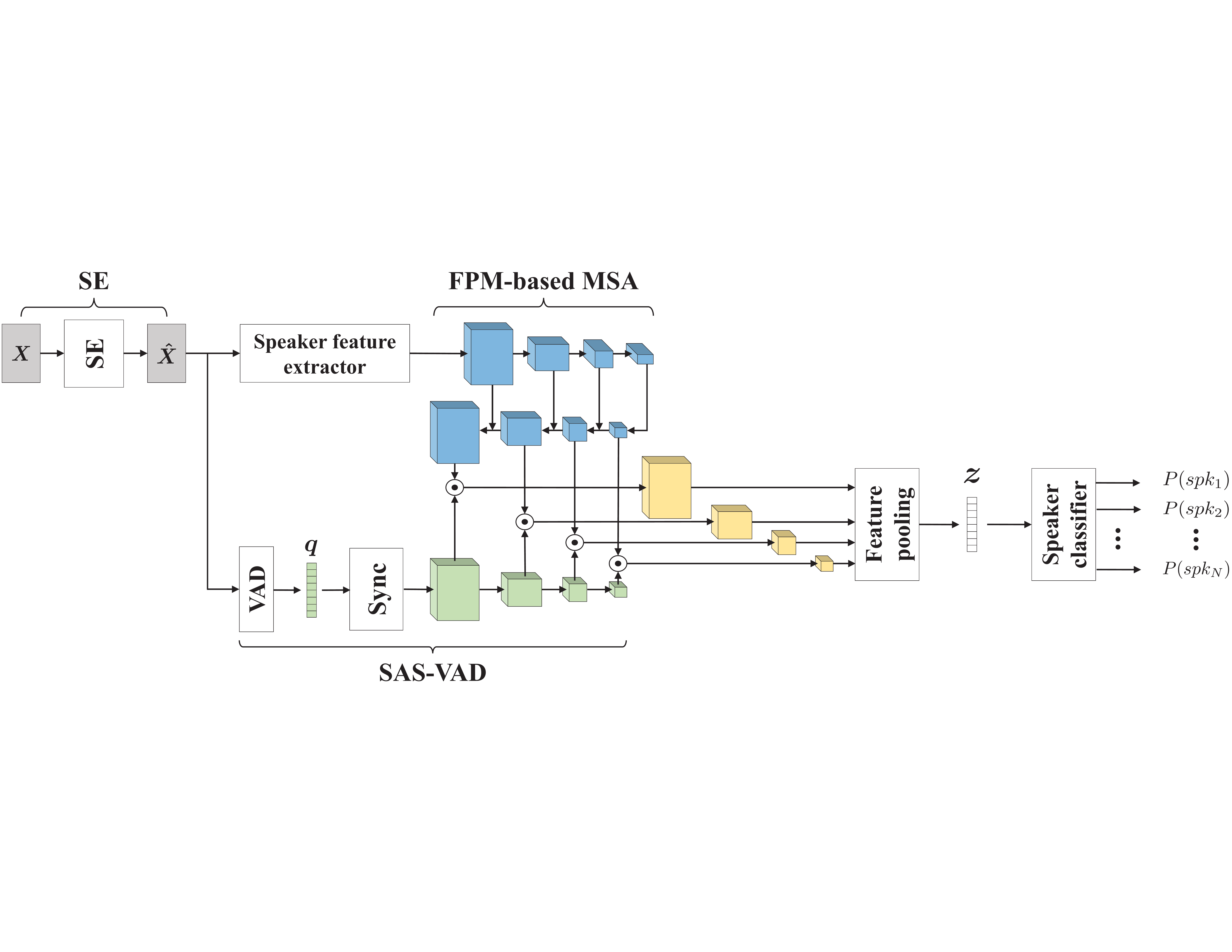}
   {\textbf{Illustration of the proposed integrated model combining speech enhancement (SE), speaker verification, and VAD.}\label{fig:total_systems}}

\subsection{Self-adaptive VAD}
\label{sec:self-adaptive_VAD}
In general, we train SV and VAD networks on different datasets. This domain mismatch causes performance degradation of VAD when we apply the VAD for SV. To reduce the domain mismatch, especially in the soft VAD, we propose to use two unsupervised domain adaptation (DA) techniques, which are speech-posterior-based DA (SP-DA) and joint-learning-based DA (JL-DA). 

In SP-DA, a pre-trained VAD network is fine-tuned on SV data, thereby requiring VAD labels for the SV data.
However, obtaining VAD labels for the SV data is costly and time-consuming in general. 
The reason is as follows. 
There are mainly three ways to generate VAD labels.
The first way is to use human labeling that produces labels by a human expert manually. 
The second one is to use forced-alignment automatic speech recognition (ASR) \cite{Kraljevski2015}. 
The third one is to apply unsupervised VAD to the clean data and use the results as the labels of the corresponding noisy data \cite{Zhang2016, Jung2017, Jung2018}.
Note that the last method requires parallel clean and noisy data.
In general, it is difficult to apply all three methods to the SV data.
Since most SV datasets are large-scale, it is difficult to use the first method.
Moreover, it is difficult to use the second and third methods because most SV datasets do not consist of clean data with which ASR or unsupervised VAD performs almost perfectly.
Therefore, we assume that there are no VAD labels for the SV data and propose to use an unsupervised domain adaptation method where the VAD network itself generates ``reliable'' VAD labels for the unlabeled SV data.
We achieve this by repeating the following steps: (1) We threshold speech posteriors, which are the outputs of the VAD network, to generate ``reliable'' labels. 
(2) We fine-tune the VAD network with the labeled data generated by the VAD network itself.
This is why we call this approach \textit{self-adaptive VAD}.

In JL-DA, we integrate the VAD network into the SV network using the soft VAD. Then, the gradient of the loss of the SV network is backpropagated through the VAD network. 
Since the VAD network is partly guided by the SV loss, it would be able to assign higher posterior probabilities for frames which are more useful for the SV task. 
The self-adaptive VAD is conducted by combining SP-DA and JL-DA.

The overall procedure of SAS-VAD is given in Algorithm \ref{alg:SSVAD}. 
Suppose $D$ is a SV dataset, which has $U$ utterances in total.
$\mathcal{X}^v_i$ and $\mathcal{X}^s_i$ are a set of acoustic feature vectors of the $i$-th utterance, which are extracted for VAD and SV, respectively:
\begin{equation}
\mathcal{X}^v_i = \{\boldsymbol{x}^v_{i,1},\dotsm,\boldsymbol{x}^v_{i,T_i}\}\,,
\end{equation}
\begin{equation}
\mathcal{X}^s_i = \{\boldsymbol{x}^s_{i,1},\dotsm,\boldsymbol{x}^s_{i,T_i}\}\,,
\end{equation}
where $\boldsymbol{x}^v_{i,t}$ and $\boldsymbol{x}^s_{i,t}$ are the feature vectors of the $t$-th frame in $\mathcal{X}^v_i$ and $\mathcal{X}^s_i$, respectively. 
Here, $T_i$ is the total number of frames in the $i$-th utterance and the superscripts $v$ and $s$ denote VAD and SV, respectively.
Both acoustic features can be different types, but for simplicity, we use the same features for both tasks.
$y^s_i$ is the speaker label for the $i$-th utterance.
We assume that we do not have VAD labels as it is usually difficult to obtain them for SV data, as mentioned above.

In SP-DA, we generate a speech posterior vector $\boldsymbol{q}_i$ in the $i$-th utterance using the pre-trained VAD.
Each speech posterior $q_{i,t}$ is compared with the predefined threshold $\delta$ of 0.7, where $q_{i,t}$ corresponds to the $t$-th frame.
If $q_{i,t}$ is larger than $\delta$, we assume that the $t$-th frame can be reliably-labeled as a speech frame.
On the other hand, if $1-q_{i,t}$, the non-speech posterior of the $t$-th frame, is larger than $\delta$, the frame is regarded as a non-speech frame.
This operation is denoted by $\mathcal{F}(\boldsymbol{q}_i, \delta)$ and generates a set of feature vectors $\boldsymbol{\bar{X}}^{v}_i$ and corresponding VAD labels $\boldsymbol{\bar{Y}}^{v}_i$. 
After that, the VAD network is fine-tuned using the obtained labeled data $\{\boldsymbol{\bar{X}}^{v}_i,\boldsymbol{\bar{Y}}^{v}_i\}$ by minimizing the loss function $\mathcal{L}_{SP}$. In \cite{Jung2019VAD}, the cross-entropy loss function is used as $\mathcal{L}_{SP}$.

In this work, instead of the cross-entropy loss, we apply focal loss \cite{Lin2017} to handle speech/non-speech class imbalance in SP-DA. 
The class imbalance is a common problem in training the VAD network because, in many cases, there is a significant mismatch between the number of speech and non-speech frames in an utterance (usually the former is larger than the latter). It has been shown that this class imbalance in training can degrade the performance of deep learning-based classifiers in various domains \cite{Buda2018}. 
To address the problem, many VAD studies insert silence at the beginning and end of each utterance to increase the ratio of non-speech frames \cite{Ghosh2011, Zhang2016, Jung2017, Jung2018}.
Unlike this heuristic approach, in \cite{Lee2020}, we proposed to use the focal loss, which was originally designed to address class imbalance in object detection task. We demonstrated that the focal loss is useful for dealing with class imbalance in the VAD and improves the performance in various class imbalance cases.
The focal loss is defined as below:
\begin{equation}\label{eq:focal_loss}
FL(p_t) = -(1-p_t)^{\gamma}\log(p_t)\,,
\end{equation}
where $(1-p_t)^\gamma$ is a modulating factor to focus training on a rare class, which is multiplied to the cross entropy loss, and $\gamma$ is a tunable \textit{focusing} parameter. Here, $p_t$ is defined as below:
\begin{equation}
p_t = \begin{cases} p & \text{if $y=1$} \\ 1-p & \text{otherwise}\,, \end{cases}
\end{equation}
where $y \in \{\pm1\}$ is the ground-truth class and $p \in [0,1]$ is the VAD's estimated probability for the speech class (i.e., the class with label $y=1$).
In the SP-DA, there is also the class imbalance problem as in the typical VAD training.
Therefore, we decide to apply the focal loss to the SP-DA.

In JL-DA, the loss $\mathcal{L}_{JL}$, which is the loss of the SV model $\mathcal{SV}$, is calculated using $\boldsymbol{X}^{s}_i$ (a fixed-length segment of 200 frames) and the corresponding label $y^s_i$ for the $i$-th utterance.
The gradients of $\mathcal{L}_{JL}$ are backpropagated through the $\mathcal{VAD}$ and $\mathcal{SV}$, respectively. 
The self-adaptive VAD is then conducted by combining the two losses as follows:
\begin{equation}\label{eq:twolosses}
\mathcal{L}_{v} = \mathcal{L}_{JL}+\lambda\mathcal{L}_{SP}\,,
\end{equation} 
where $\mathcal{L}_{v}$ is the total loss and $\lambda$ is the loss weight for $\mathcal{L}_{SP}$. We indicate the combination of the soft VAD and the self-adaptive VAD as \textit{self-adaptive soft VAD}.

\section{Proposed integrated model}
\label{sec:integrated_method}
Fig. \ref{fig:total_systems} illustrates the proposed integrated model which consists of three models: (1) speech enhancement (SE) model using a masking network, (2) speaker verification (SV) model using FPM-based MSA, and (3) voice activity detection (VAD) model using modified self-adaptive soft VAD (SAS-VAD). 
The overall procedure of the proposed approach is described in the following paragraph.

The masking network in the SE model estimates the ratio mask $\bm{M}$ and the resulting mask is multiplied with the corrupted feature matrix $\bm{X}$ element-wise to produce the enhanced feature matrix $\bm{\hat{X}}$, as explained in Section \ref{sec:speech-enhancement}. 
Then, $\bm{\hat{X}}$ is fed into both SV and VAD models. 
In the SV model, multi-scale feature maps are extracted by a speaker feature extractor and enhanced by the FPM, as explained in Section \ref{sec:FPM}. 
In the VAD model, a speech posterior vector $\bm{q}$ is produced and fed into the synchronizer (denoted as Sync in the figure). After that, we obtain four reduced versions of $\bm{q}$ and perform the soft VAD by multiplying the enhanced feature maps by their corresponding speech posterior vectors, as explained in Section \ref{sec:soft-VAD}. 
The resulting feature maps are colored in yellow. Then, the feature maps are converted into the speaker embedding $\bm{z}$ by a feature pooling layer, which consists of SAP and FC layers. Finally, $\bm{z}$ is fed into a speaker classifier. 
All the networks are jointly trained in an end-to-end manner to classify training speakers using softmax loss.
Furthermore, the VAD model is adapted to the SV data by using self-adaptive VAD with focal loss.

Our end-to-end approach has two advantages over the conventional approach using separately pre-trained models. 
First, since the VAD and SE models are optimized to minimize the SV loss, they do not require labels for training, which are difficult to obtain for most SV datasets. 
In Section \ref{sec:self-adaptive_VAD}, we explained why it is difficult to obtain VAD labels.
In the case of the SE, it is also difficult to obtain labels for the same reason.
The SE model is usually optimized by minimizing the mean square error (MSE) between the enhanced and clean speech features \cite{Meng2018}. 
That is, we use the clean features as the label of its corresponding input noisy features.
As already explained, obtaining parallel clean and noisy data is difficult for the SV data.
The second advantage is that both VAD and SE models are guided by the SV loss to generate outputs which are more suitable and useful for the SV task. 
In the case of the VAD, this approach is called joint-learning-based domain adaptation (see Section \ref{sec:self-adaptive_VAD}). 
In the experiments, we will show that our approach performs better than using a pre-trained VAD model without adaptation.

\section{EXPERIMENTAL SETUP}
\label{sec:DB_systems}

\subsection{Datasets}
For speaker verification, we used two datasets: Korean indoor (KID) \cite{Suh2017} and VoxCeleb \cite{Nagrani2017}.
The KID dataset is a text-independent dataset consisting of reverberated speech and noise. It was collected at a distance of 3 m from the source in an indoor environment, which is a simulated living room with the reverberation time (RT60) of 0.23 s. 
Compared to the corrupted data generated by simply adding prerecorded noise to clean speech collected independently of each other, the corrupted data from the KID dataset is much closer to natural data since both speech and noise were collected in the same room with the same microphone.

We used the same data setup as \cite{Jung2019VAD}.
There are a total of 550 speakers for training and validation.
For each utterance, the noise was randomly selected from three types of noise, i.e., air conditioner, smartphone ringtone, and TV, and added to the reverberated speech at randomly selected signal-to-noise ratios (SNRs) between 0 and 10 dB, resulting in 200 utterances per speaker. 
Here, utterances of randomly selected 20 speakers were used for validation and the rest of them were used for training. 
The utterances of other 105 speakers were used for testing. 
We inserted silence at the beginning and end of the utterance to simulate more realistic conditions where the need for robust VAD is higher, which will be explained later.
After inserting silences, the noise was randomly selected from three types of noise, i.e., refrigerator, babble, and music, and added to the reverberated speech at randomly selected SNRs of 0, 5, and 10 dB, resulting in 24 utterances per speaker. 12 utterances were sampled as the enrollment data for each speaker. Other than 12 enrolled utterances, we sampled 12 utterances each from the same and different speakers. 
A total of 30 k trials were generated for testing.

VoxCeleb is a dataset for large scale text-independent speaker verification containing 1,250 speakers, and the dataset is split into development (dev) and test sets.
There are no overlapping speakers between them.
The utterances were extracted from YouTube videos, which are corrupted by real-world noise.
For training and testing, we used the similar data augmentation strategy as in \cite{Shon2019}.
That is, we augmented the dev set with additive noises from MUSAN dataset \cite{Snyder2015} and reverberation from RIR dataset \cite{Ko2017}. 
For simplicity, we denote the combination of both datasets as noise dataset, which consists of four types of acoustic distortions: babble, music, noise, and reverberation.
We split the noise dataset into two disjoint subsets and used each of them to augment the dev and test set, respectively. In both the dev and test sets, we used the same noise types, but different noise samples. 
The dev set contains 148,642 utterances from 1,211 speakers. 
We corrupted each utterance at SNR levels varying from 0 to 20 dB. 
The resulting augmented set has the same amount of data as the original dev set.
The test set contains 4,715 utterances from 40 speakers and has 37,720 verification trials in total, including 18,860 trials for each positive and negative trial. 
As in the case of the KID dataset, we inserted silence at the beginning and end of the test utterance to simulate the environments where the need for robust VAD is higher. 
After inserting silences, we added the four types of noise at three levels of SNRs: 0, 5, and 10 dB, respectively. Thus, we totally generated 12 corrupted test sets.
Each corrupted test set has the same amount of data as the original test set.

To evaluate the performance on short speech segments, we modified the original test set to construct test sets of four different short durations (1, 2, 3, and 4 s) before inserting silence.
For the KID dataset, if the length of the utterance was less than the given duration, we concatenated two or more utterances until the total length reached the given duration.
For the VoxCeleb dataset, if the length of the utterance was less than the given duration, the entire utterance was used.
To sum up, we first constructed the test sets of different short durations. Then, silence was appended at the beginning and end of each utterance. Finally, we added noise or reverberation to the utterance.
Note that the generated test data contains short speech segments and long non-speech segments degraded by noise and reverberation.
In this paper, we use the notation `S$x$-N$y$' to denote a condition in which $x$ corresponds to the length of the speech segment and $y$ corresponds to the length of the non-speech segments (i.e., padded silence). For example, `S4-N2' means that the length of the speech segment is 4 s and 1 s of silence is padded at the beginning and end of the utterance (i.e., in total 2 s of non-speech segments). 

\begin{table}[t]
\centering
\begin{scriptsize}
\renewcommand{\arraystretch}{1}
\caption{\textbf{ResNet architectures for deep speaker embedding. The shape of a residual block is shown inside the brackets and the number of stacked blocks on a stage is shown outside the brackets. $d$ : dimension of acoustic features.}}
\label{SR_architecture}
\vspace{-0.1cm}
\begin{tabular}{cccc}
\cline{1-4} 
\addlinespace[0.3ex]
\textbf{Layer} &  \textbf{1D-Res34}    & \textbf{2D-Res34} & \textbf{Stage} \\   
\addlinespace[0.3ex] \cline{1-4} \addlinespace[0.5ex]
\cline{1-4}
\addlinespace[0.6ex]
conv1      & $7 \times d, 32$, stride $1$     & $7 \times 7, 32$, stride $1$   & -                                           \\ 
\addlinespace[0.7ex]
\cline{1-4}
\addlinespace[0.7ex]
conv2\_x      & $\begin{bmatrix} 3 \times 1, 32 \\ 3 \times 1, 32  \end{bmatrix} \times 3$   & $\begin{bmatrix} 3 \times 3, 32 \\ 3 \times 3, 32  \end{bmatrix} \times 3$  & 1 \\ 
\addlinespace[0.7ex]
\cline{1-4}
\addlinespace[0.7ex]
conv3\_x        & $\begin{bmatrix} 3 \times 1, 64 \\ 3 \times 1, 64  \end{bmatrix} \times 4$   & $\begin{bmatrix} 3 \times 3, 64 \\ 3 \times 3, 64  \end{bmatrix} \times 4$ & 2 \\ 
\addlinespace[0.7ex]
\cline{1-4}
\addlinespace[0.7ex]
conv4\_x       & $\begin{bmatrix} 3 \times 1, 128 \\ 3 \times 1, 128  \end{bmatrix} \times 6$  & $\begin{bmatrix} 3 \times 3, 128 \\ 3 \times 3, 128  \end{bmatrix} \times 6$ & 3 \\ 
\addlinespace[0.7ex]
\cline{1-4}
\addlinespace[0.7ex]
conv5\_x        & $\begin{bmatrix} 3 \times 1, 256 \\ 3 \times 1, 256  \end{bmatrix} \times 3$  & $\begin{bmatrix} 3 \times 3, 256 \\ 3 \times 3, 256  \end{bmatrix} \times 3$ & 4  \\ 
\addlinespace[0.7ex]
\cline{1-4}
\addlinespace[0.5ex]
\cline{1-4}
\addlinespace[0.3ex]
global pooling       & \multicolumn{2}{c}{GAP, SAP, or ASP} & -  \\ 
\addlinespace[0.3ex]
\cline{1-4}
\addlinespace[0.3ex]
FC1 &  \multicolumn{2}{c}{$k \times 128$} & - \\
\addlinespace[0.3ex]
\cline{1-4}
\addlinespace[0.5ex]
\cline{1-4}
\addlinespace[0.3ex]
FC2 & \multicolumn{2}{c}{$128\,\times$ \# speakers} & -   \\
\cline{1-4}

\end{tabular}
\end{scriptsize}
\end{table}

For VAD, we used the same data setup as in \cite{Jung2018}, where noisy data are generated by corrupting the clean utterances of the Aurora4 \cite{Pearce2002} with noise.
For training, we inserted 2 s of silence at the beginning and end of the utterance to address the speech/non-speech class imbalance. 
Then, we added 100 types of noise\footnote{web.cse.ohio-state.edu/pnl/corpus/HuNonspeech/HuCorpus.html.\\100 types of noise (the total number of files is indicated in parentheses) : Crowd (17), Alarm and siren (14), Water sound (14), Machine (12), Animal sound (9), Wind (9), Bell (4), Cough (3), Laugh (3), Traffic and car (3), Door moving (2), Yawn (2), Clap (1), Click (1), Cry (1), Footsteps (1), Phone dialing (1), Shower (1), Snore (1), Tooth brushing (1).} \cite{Hu2010}
to the clean data at randomly selected SNRs of -5, 0, 5, 10, 15, and 20 dB. 
For testing, all the 330 utterances of the Aurora4 clean test set were used. We added five unseen noises (babble, car, street, factory, and F16 cockpit) in the NOISEX-92 noise database \cite{Varga1993} at three low SNR levels: -5, 0, and 5 dB. 
We applied Sohn VAD \cite{Sohn1999} to the clean data and used the results as VAD labels of the corresponding noisy data as in \cite{Zhang2016, Jung2018}.

\subsection{Model architectures}
We built the baseline \textit{i}-vector/PLDA system using the Kaldi toolkit \cite{Povey2011}.
We extracted 20-dimensional Mel-frequency cepstral coefficients (MFCC) features from the utterances with a 25 ms Hamming window. 
Delta and acceleration were appended to generate 60 dimensional features. 
An energy-based VAD was used to select features corresponding to speech frames. 
The UBM contains 2048 Gaussian mixtures and the dimension of \textit{i}-vector is set to 600.
Prior to PLDA scoring, \textit{i}-vectors were centered and length normalized.

For deep speaker embedding learning, we built three different architectures (TDNN, 1D-ResNet34, and 2D-ResNet34) using PyTorch \cite{Paszke2017}. 
We followed the same TDNN architecture as in \cite{Okabe2018}. 
The detailed architectures of 1D-ResNet and 2D-ResNet are described in Table \ref{SR_architecture}, where both networks have 34 layers. 
The 1D-ResNet is based on the feature extractor of \cite{Bhattacharya2019}. 
The first conv. layer utilizes a $3 \times d$ filter, where $d$ is the dimension of the acoustic features. 
Other conv. layers perform 1-dimensional convolution operations along the time axis with filter size 3. 
The 2D-ResNet is based on the feature extractor of \cite{Jung2020}. 
For both networks, conv. layers constitute a speaker feature extractor, and the following feature pooling layer (i.e., global pooling and FC1) converts the output feature maps to a fixed-dimensional speaker embedding. 
When we do not use MSA, $k$ is set to 256 which is the number of filters of the last conv. layer. 
When we use MSA, $k$ is the sum of the number of filters of all selected conv. layers.
The final FC layer (i.e., FC2) is fed to a softmax function to produce a probability distribution over all speakers in the training set. 
We extracted 128-dimensional speaker embeddings from FC1 after training.

\begin{table}[t]
\centering
\begin{footnotesize}
\renewcommand{\arraystretch}{1}
\caption{\textbf{Detailed network configurations of the DNN, LSTM, and CLDNN-based VADs.}}
\label{VAD_architecture}
\vspace{-0.1cm}
\begin{tabular}{cccc}
\hline
\textbf{Model}        & \textbf{DNN}         & \textbf{LSTM}        & \textbf{CLDNN}       \\ \hline\hline
\textbf{CNN layer}  & \multicolumn{1}{l}{} & \multicolumn{1}{l}{} & \multicolumn{1}{l}{} \\ \hline
\# filter outputs          &     -                 &    -                  & 42                   \\
filter size (time $\times$ freq)  &     -                 &     -                & 1$\times$8                  \\
pooling size (time $\times$ freq) &     -                 &     -                 & 1$\times$3                  \\ \hline
\textbf{LSTM layers}       &                      &                      &                      \\ \hline
\# LSTM hidden layers      &     -                 & 3                    & 1                    \\
\# hidden units per layer  &      -                & 42                   & 42                   \\ \hline
\textbf{Fully-connected layers}        &                      &                      &                      \\ \hline
\# Fully-connected hidden layers       & 2                    &  -                     & 1                    \\
\# hidden units per layer  & 64                   &  -                   & 42                   \\ \hline\hline
\# parameters & 4.97k                & 4.89k                & 4.75k             \\ \hline
\end{tabular}
\end{footnotesize}
\end{table}

\begin{table*}[t]
\centering
\begin{footnotesize}
\renewcommand{\arraystretch}{1.1}
\caption{\textbf{EERs (\%) of three feature extractors (TDNN, 1D-Res34, and 2D-Res34) using two acoustic features (Fbank64 and Spec160), on KID and original VoxCeleb datasets. Four test sets of different durations are evaluated on 4 s enrollment set. In all tables, we highlight the enrollment condition in bold.}}
\label{TDNN_ResNet_comparison}
\vspace{-0.1cm}
\setlength\tabcolsep{5pt}
\begin{tabular}{c||cccc|cccc||cccc|cccc}
\hline
\multirow{3}{*}{System}    & \multicolumn{8}{c||}{KID dataset}                            & \multicolumn{8}{c}{Original VoxCeleb}                        \\ \cline{2-17} 
                           & \multicolumn{4}{c|}{w/o MSA}   & \multicolumn{4}{c||}{w/ MSA}  & \multicolumn{4}{c|}{w/o MSA}   & \multicolumn{4}{c}{w/ MSA}  \\ \cline{2-17} 
                           & 1 s   & 2 s   & 3 s   & \textbf{4 s}   & 1 s   & 2 s   & 3 s   & \textbf{4 s}   & 1 s   & 2 s   & 3 s   & \textbf{4 s}   & 1 s   & 2 s   & 3 s   & \textbf{4 s}   \\ \hline
\textit{i}-vector/PLDA     & 24.50 & 15.26 & 11.83 & 9.98 & - & - & - & - & 18.49 & 11.96 & 8.91  & 7.46 & - & - & - & - \\ \hline 
TDNN-Fbank64        & 19.00 & 11.23 & 9.07 & 7.87 & 18.18 & 11.19 & 8.88 & 7.69 & 12.27 & 8.44 & 7.30 & 6.79 & 12.53 & 8.66 & 7.51 & 7.00 \\ 
TDNN-Spec160        &   17.35    &  10.65     &     8.40  &     7.01  &     17.31  &    10.65   &  8.24     &  6.69     &   13.67    & 9.58      &  7.97     &  7.45     &   13.81    &     9.77  & 8.13  & 7.47  \\ \hline
1D-Res34-Fbank64 &   18.10    &    11.68   &  9.67     &  8.65     &  15.78     &  9.41     & 7.61      &  6.70     &     12.73  & 9.19      &  7.97  &    7.50   &     10.93 & 8.10  &  7.03  &   6.61    \\
1D-Res34-Spec160 &  17.62     &    10.17   &  7.65     &  6.31     &  16.92     &  9.42     &  7.16     &  5.81     &   13.23    &     9.34  & 7.81      & 7.32  & 11.23      &     8.30  &    6.95   &     6.49  \\ \hline
2D-Res34-Fbank64 &   16.28    &    10.19   &  8.10     &  7.22     &   15.21    &  9.46     &  7.34     &  6.37  &     9.45  &  6.83     &   5.75    &   5.40    &  \textbf{8.25}     &  \textbf{5.75}     &  \textbf{4.93}     &  \textbf{4.55}     \\
2D-Res34-Spec160 &    14.13   &    7.85   &    5.81   &    5.08   &  \textbf{13.17}     &  \textbf{7.07}     &  \textbf{5.15}     & \textbf{4.52}      & 9.95  &  7.19  &  6.17 &  5.79 &   8.54    &  6.22     &  5.43     & 5.06      \\ \hline
\end{tabular}
\end{footnotesize}
\end{table*}

We used three types of deep architectures for VAD: DNN, LSTM, and CLDNN \cite{Zazo2016}, as shown in Table \ref{VAD_architecture}.  
The DNN model is a fully-connected neural network with 2 hidden layers and 64 hidden units per layer.
A ReLU function is used for each hidden layer. 
The LSTM model uses 3 unidirectional LSTM layers with 42 hidden units per layer. The LSTM is unrolled for 50 time steps for training with truncated backpropagation through time (BPTT). 
The CLDNN model uses DNN, CNN, and LSTM layers in a unified framework. The first layer consists of a conv. layer with kernel size $1 \times 8$ in time $\times$ frequency. After the conv. operation, we apply non-overlapping max pooling along the frequency axis, with pooling size 3. 
The output of the conv. layer is passed to one LSTM layer, and then to one fully-connected layer. 
The output layer of all three models uses a sigmoid activation function to predict speech.
Note that, we established a fair comparison among three models with a comparable number of total parameters ($\approx$ 5 k).

\begin{table}[t]
\centering
\begin{footnotesize}
\renewcommand{\arraystretch}{1.1}
\caption{\textbf{Hyperparameters of self-adaptive soft VAD.}}
\label{lambda_gamma}
\vspace{-0.1cm}
\begin{tabular}{c|c|c|c|c}
\hline
\multirow{2}{*}{\begin{tabular}[c]{@{}c@{}}Hyper\\ parameter\end{tabular}} & \multicolumn{2}{c|}{KID dataset} & \multicolumn{2}{c}{VoxCeleb} \\ \cline{2-5} 
                                                                           & w/ SE          & w/o SE          & w/ SE         & w/o SE        \\ \hline
$\lambda$ in (\ref{eq:twolosses})                                                                    & 2              & 4               &   2            &  5             \\ \hline
$\gamma$ in (\ref{eq:focal_loss})                                                                     & 0.5               & 0.5                &    0.1           & 0.1              \\ \hline
\end{tabular}
\end{footnotesize}
\end{table}

\subsection{Implementation details}
\label{sec:implementation_details}
We extracted two types of acoustic features: 64-dimensional log Mel-filterbank features (Fbank64) and 160-dimensional log spectrogram (Spec160) with a 25 ms Hamming window and 50\% overlap using a 512-point FFT. In the case of Spec160, we reserve 0-5 k range to make the spectrogram with a dimension of 160. As we mentioned in Section \ref{sec:self-adaptive_VAD}, we used the same features for both SV and VAD. When using Spec160, we modified the ResNet-based feature extractor slightly to reduce the frequency dimension of the feature because the frequency dimension of Spec160 is much larger than Fbank64. 
To do so, we changed the stride of conv1 layer to $1 \times 2$ and added an additional $2 \times 2$ max-pooling layer with stride $1 \times 2$, thus reducing the frequency dimension by a factor of four in the lowest layer.

For SV, we applied mean normalization over an input segment. In training, the input size is $d \times 200$ for 2 s segment, where $d$ is 64 for Fbank64 or 160 for Spec160. 
In testing, the entire utterance was evaluated at once. 
We report the equal error rate (EER) in \%. Verification trials were scored using cosine distance. In the case of the VoxCeleb, we have totally 12 test sets for each `S$x$-N$y$'. Therefore, we report the average EER across all 12 test sets for each `S$x$-N$y$'.
All models were optimized using stochastic gradient descent with momentum 0.9. The weight decay parameter is 0.0001, and the batch size is 64.
We used the same learning rate schedule as in \cite{Jung2020} with the
initial learning rate of 0.1.
When applying MSA, the parameters of the SAP layers are not shared by all ResNet stages.

For VAD, the acoustic features were normalized based on the global mean and standard deviation of the whole training set. 
For the DNN-based VAD, all acoustic feature vectors were augmented with neighboring frames within a context window of size 11 (i.e., total 11 frames) and fed into the network. 
We used the Adam optimizer with the initial learning rate of $10^{-5}$ and a batch size of 512. 
For self-adaptive soft VAD, we used the same learning rate schedule as in the SV with the initial learning rate of $10^{-7}$.
The hyperparameters $\lambda$ in (\ref{eq:twolosses}) and $\gamma$ in (\ref{eq:focal_loss}) are given in Table \ref{lambda_gamma}.

\begin{figure*}[!t]
    \centering
    \begin{footnotesize}
    \renewcommand{\tabcolsep}{7mm}
    \renewcommand{\arraystretch}{0.4}
        \begin{tabular}{ccc}
            \includegraphics[trim=0.0cm 0.1cm 0.0cm 0.0cm, clip=true, width=3.96cm]{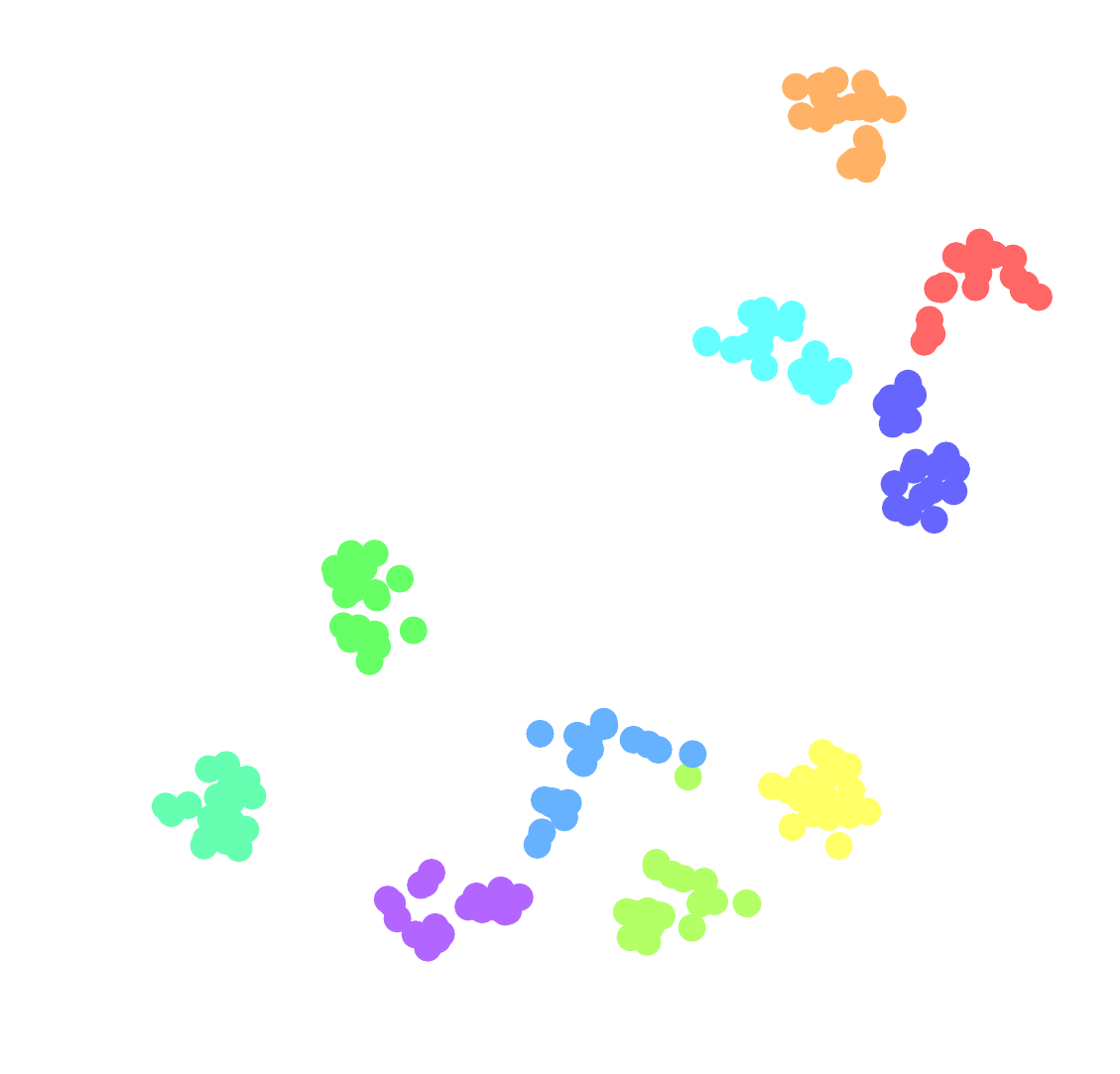} &
            \includegraphics[trim=0.0cm 0.1cm 0.0cm 0.0cm, clip=true, width=3.96cm]{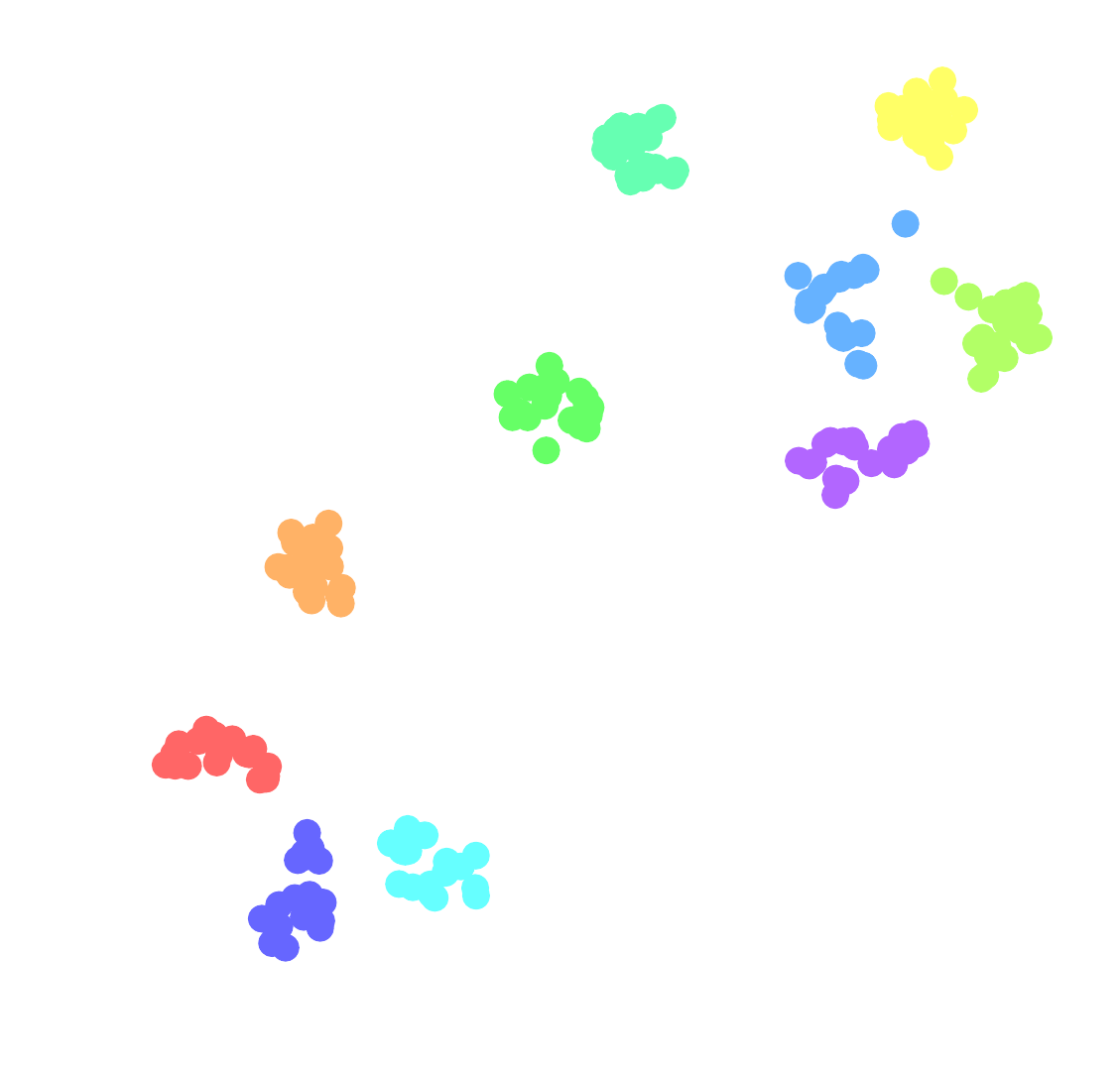} &
            \includegraphics[trim=0.0cm 0.1cm 0.0cm 0.0cm, clip=true, width=3.96cm]{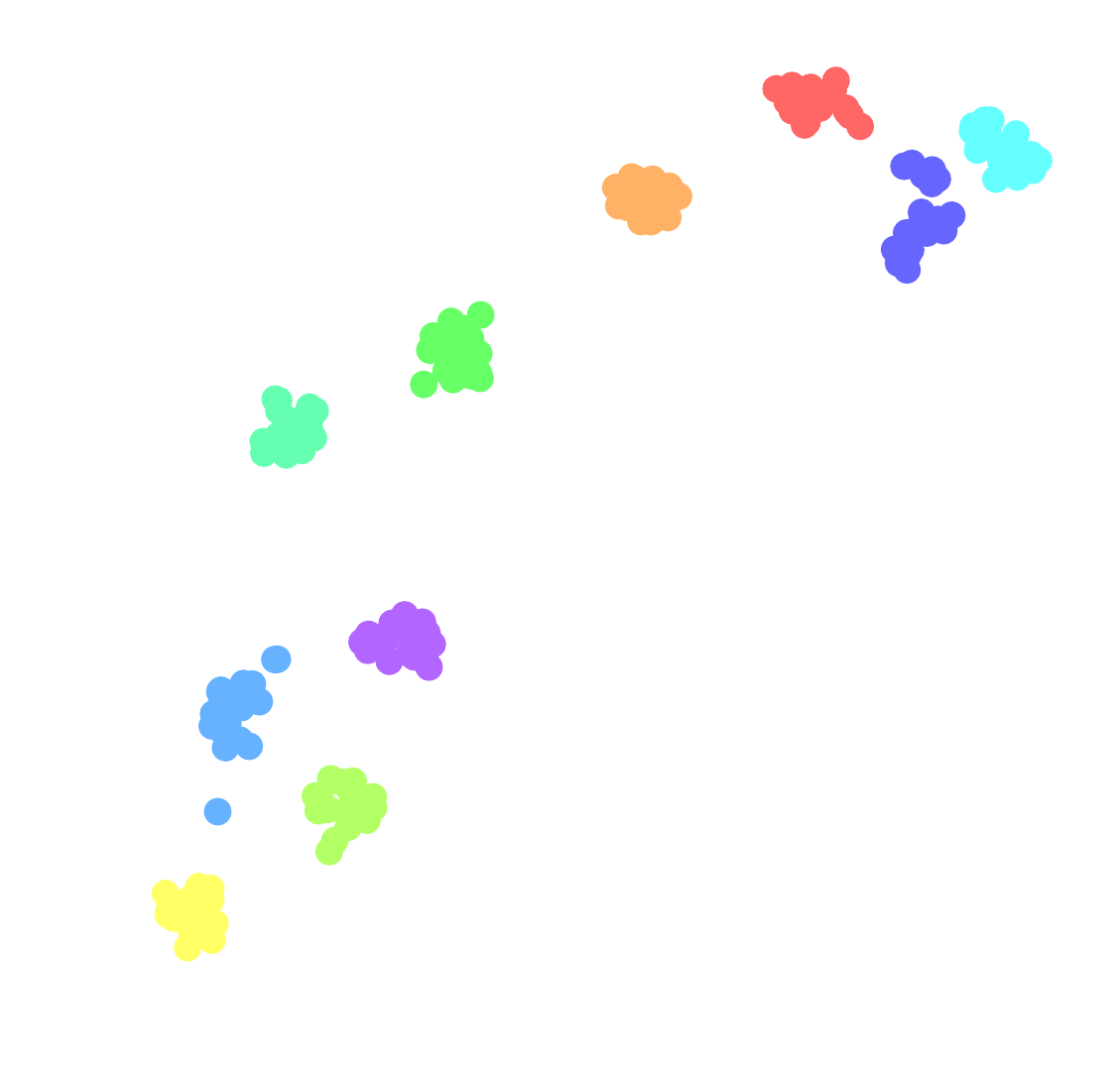} 
            \\
            (a) w/o MSA & (b) MSA w/o FPM & (c) MSA w/ FPM
        \end{tabular}
    \end{footnotesize}
\vspace{-0.1cm}
\caption{\textbf{t-SNE visualization of speaker embeddings extracted from the KID dataset, where each color corresponds to a different speaker.}}
\vspace{-0.35cm}
\label{fig:tsne}
\end{figure*}

\section{EXPERIMENTAL RESULTS}
\label{sec:results}

\subsection{Impact of multi-scale aggregation}
To verify the effectiveness of MSA, we compare the EERs between the deep speaker embedding systems without MSA (w/o MSA) and with MSA (w/ MSA) for the two datasets: KID and original VoxCeleb. 
Here, w/o and w/ MSA are correspond to the approaches of Fig. \ref{fig:PFH_FPN}(a) and (b), respectively.
In this experiment, we only used the original VoxCeleb dataset without augmentation for both training and testing.
To evaulate the performance on short speech segments, we performed experiments on four test sets of different durations: 1, 2, 3, and 4 s. We set the duration of enrollment data to 4 s. 

Table \ref{TDNN_ResNet_comparison} presents the experimental results of three feature extractors (TDNN, 1D-Res34, and 2D-Res34) using two acoustic features (Fbank64 and Spec160). 
In all cases, we use the softmax loss for training and the GAP for feature pooling. 
There are five layers in the TDNN, and we extract feature maps from the highest four layers when applying the MSA.
The EERs of the baseline \textit{i}-vector/PLDA system are also reported.
When we compare the \textit{i}-vector/PLDA system with other deep speaker embedding systems, we can obviously see that the latter outperforms the former in all the cases. As the test duration decreases, the performance gap becomes larger. 
This implies that the deep speaker embedding is more effective than the \textit{i}-vector/PLDA baseline to deal with short speech segments.
Especially, for the KID dataset, 2D-Res34-Spec160 w/ MSA achieves a relative improvement of 46.24\% over \textit{i}-vector/PLDA on the 1 s condition. 
In the case of the TDNN-based feature extractor, we can observe that the MSA does not bring a significant improvement, while the MSA is clearly beneficial for both ResNet-based feature extractors on all test conditions.
Even in some cases, the MSA degrades the performance of the TDNN-based system.
That is, even though we extract feature maps from four different layers, using these features is only useful for the ResNet-based feature extractors.
This is because the TDNN does not have any local pooling layers, thus all the feature maps from different layers have the same scale (i.e., operating at the frame-level). Strictly speaking, the use of the term ``MSA'' is not appropriate for the TDNN case because multi-scale features are not generated in the TDNN, even though they are extracted from different layers. 
These results suggest that using multi-scale features improves the deep speaker embedding learning on various short-duration conditions.

\begin{table}[t]
\centering
\begin{footnotesize}
\renewcommand{\arraystretch}{1.0}
\caption{\textbf{Ablation results of FPM on the KID dataset (EER \%).}}
\label{ablation_FPM_KID}
\vspace{-0.2cm}
\setlength\tabcolsep{5pt}
\begin{tabular}{c|c|cc||cccc}
\hline
System                        & FM & LAT & TD & 1 s & 2 s & 3 s & \textbf{4 s} \\ \hline
w/o MSA                       & $C_5$  &    $\times$    &    $\times$     &    16.28 & 10.19   & 8.10    & 7.22    \\ 
MSA w/o FPM                       & \{$C_k$\}  &   $\times$     &    $\times$     &    15.21 & 9.46   & 7.34    & 6.37    \\ \hline
\textbf{MSA w/ FPM}                       & \{$P_k$\}  & \checkmark       & \checkmark        &    \textbf{14.71} & \textbf{9.02}   & \textbf{7.15}    & \textbf{6.28}    \\ \hline
w/o TD  & \{$C_k$\}  & \checkmark       &    $\times$      &  14.75    &  9.57   & 7.60 & 6.92   \\ 
w/o LAT & \{$P_k$\}  &     $\times$    & \checkmark        & \textbf{14.71}   & 9.14   & 7.32    & 6.39   \\ 
only $P_2$             & $P_2$      & \checkmark      & \checkmark        & 16.18   & 10.50   & 8.80   &     7.63 \\ \hline
\end{tabular}
\vspace{-0.2cm}
\end{footnotesize}
\end{table}

\subsection{Impact of feature pyramid module}

To identify the contribution of the individual components in the feature pyramid module (FPM), we performed ablation studies. Table \ref{ablation_FPM_KID} and \ref{ablation_FPM_VoxCeleb} show the results of the ablation studies on the KID and VoxCeleb datasets, respectively. 
From this section, we denote the combination of the original VoxCeleb and the augmented VoxCeleb as VoxCeleb, where each test was performed on 12 augmented test sets. 
For both ablation studies, we used 2D-Res34-Fbank64 as a feature extractor with softmax loss for training and GAP for feature pooling. 
In the tables, ``FM'' denotes the selected feature maps in the feature extractor. 
As explained in Section \ref{sec:FPM}, $C_k$ denotes the feature map from the bottom-up pathway (i.e., without the top-down pathway) and $P_k$ denotes the feature map from the FPM. 
Here, \{$C_k$\} and \{$P_k$\} mean that a system uses all $C_k$ and $P_k$, respectively, for $k$ = 2, 3, 4, 5.
``LAT'' and ``TD'' stand for the lateral connections and top-down pathway, respectively, which are the components of the FPM.

\begin{table}[t]
\centering
\begin{footnotesize}
\renewcommand{\arraystretch}{1.0}
\caption{\textbf{Ablation results of FPM on the VoxCeleb dataset (EER \%).}}
\label{ablation_FPM_VoxCeleb}
\vspace{-0.2cm}
\setlength\tabcolsep{5pt}
\begin{tabular}{c|c|cc||cccc}
\hline
System                        & FM & LAT & TD & 1 s & 2 s & 3 s & \textbf{4 s} \\ \hline
w/o MSA                       & $C_5$  &     $\times$   &    $\times$     &    13.77 & 10.15   & 8.59    & 8.01    \\ 
MSA w/o FPM                       &  \{$C_k$\}   &    $\times$    &    $\times$     &    13.11 & 9.63   & 8.20    & 7.68    \\ \hline
\textbf{MSA w/ FPM}                       & \{$P_k$\}  & \checkmark       & \checkmark        &    \textbf{12.42} & \textbf{9.38}   & \textbf{8.11}    & \textbf{7.62}    \\ \hline
w/o TD  & \{$C_k$\}  & \checkmark       &     $\times$     &  12.98    &  9.59   & 8.16 & 7.63   \\
w/o LAT & \{$P_k$\}  &    $\times$     & \checkmark        & 13.40   & 9.72   & 8.33    & 7.76   \\ 
only $P_2$             & $P_2$      & \checkmark      & \checkmark        & 14.49   & 10.66   & 9.01   &     8.35 \\ \hline
\end{tabular}
\vspace{-0.2cm}
\end{footnotesize}
\end{table}

In the 1st row of the tables, w/o MSA corresponds to the system using only the last feature map $C_5$, i.e., without applying multi-scale aggregation (MSA). 
In the 2nd row, MSA w/o FPM corresponds to the system using the MSA without the FPM.
In the 3rd  row, MSA w/ FPM corresponds to the system using the FPM-based MSA, which has all the components of the FPM. This system shows the best performance on all test cases for both datasets. 
We use t-SNE \cite{Maaten2008} to visualize the learned speaker embeddings in Fig. \ref{fig:tsne}(a), (b), and (c), using 4 s utterances.
Here, (a), (b), and (c) correspond to the first three rows of the table, respectively.
These figures show the distribution of speaker embeddings from 10 speakers randomly chosen from the test set of the KID dataset.
When comparing (a) and (b), we can see that using the MSA enhances intra-class compactness in that the embeddings of the same speaker are closer to each other.
Likewise, when comparing (b) and (c), we can observe that using the FPM further enhances the intra-class compactness.

The 4th, 5th, and 6th rows show the results of the ablation experiments.
The 4th row presents the result of the FPM without the top-down pathway (w/o TD). 
To validate the effectiveness of the top-down pathway, we remove the top-down pathway from the FPM.
In this architecture, the $1 \times 1$ lateral connections followed by $3 \times 3 $ convolutions are directly attached to the bottom-up pathway. 
Note that this architecture is different from w/ MSA in Table \ref{TDNN_ResNet_comparison}, in that the w/ MSA does not have the lateral connections and $3 \times 3 $ convolutions. 
For all test cases, w/o TD gives higher EERs compared to the w/ FPM. 
We conjecture that this is because there are large differences in the amount of speaker-discriminative information between different layers of the feature extractor. 
The FPM enhances the speaker-discriminative information of lower-layer feature maps, so that it can improve the performance of the MSA. 
The 5th row shows the ablation results of the FPM without the lateral connections. 
From the results, we can say that the lateral connections improve the FPM by transferring the information from the bottom-up pathway to the top-down pathway. 
The last row reveals the results of the FPM-based MSA only using $P_2$, which is the highest-resolution feature maps with the highest speaker-discriminative information. We can observe that ``only $P_2$'' has much lower performance than the proposed method. This reveals that using multi-scale feature maps is important even when the FPM is used.  

\begin{table}[t]
\centering
\begin{footnotesize}
\renewcommand{\arraystretch}{1.1}
\caption{\textbf{EERs (\%) of systems with and without speech enhancement. FPM-based MSA, softmax loss, and self-attentive pooling are used.}}
\vspace{-0.2cm}
\label{ablation_SE}
\begin{tabular}{c|c||cccc}
\hline
Dataset                                      & SE & 1 s & 2 s & 3 s & \textbf{4 s} \\ \hline
\multirow{2}{*}{KID}                           & $\times$   & 14.82   & 9.38   & 7.19   & 6.21 \\  
                                                 &  \checkmark  &  \textbf{14.17}  & \textbf{8.68}   &  \textbf{6.54}  & \textbf{5.73}    \\ \hline
\multirow{2}{*}{VoxCeleb} &   $\times$  & 13.24   & 9.75   & 8.28   & 7.78    \\  
                           &   \checkmark  & \textbf{13.10}   &  \textbf{9.44}  & \textbf{7.69}   & \textbf{7.21}    \\ \hline
\end{tabular}
\end{footnotesize}
\end{table}

\begin{figure}[!t]
    \centering
    \begin{footnotesize}
    \renewcommand{\tabcolsep}{0.5mm}
    \renewcommand{\arraystretch}{1.4}
        \begin{tabular}{cc}
            \includegraphics[trim=0.0cm 0.0cm 0.0cm 0.0cm, clip=true, width=4.3cm]{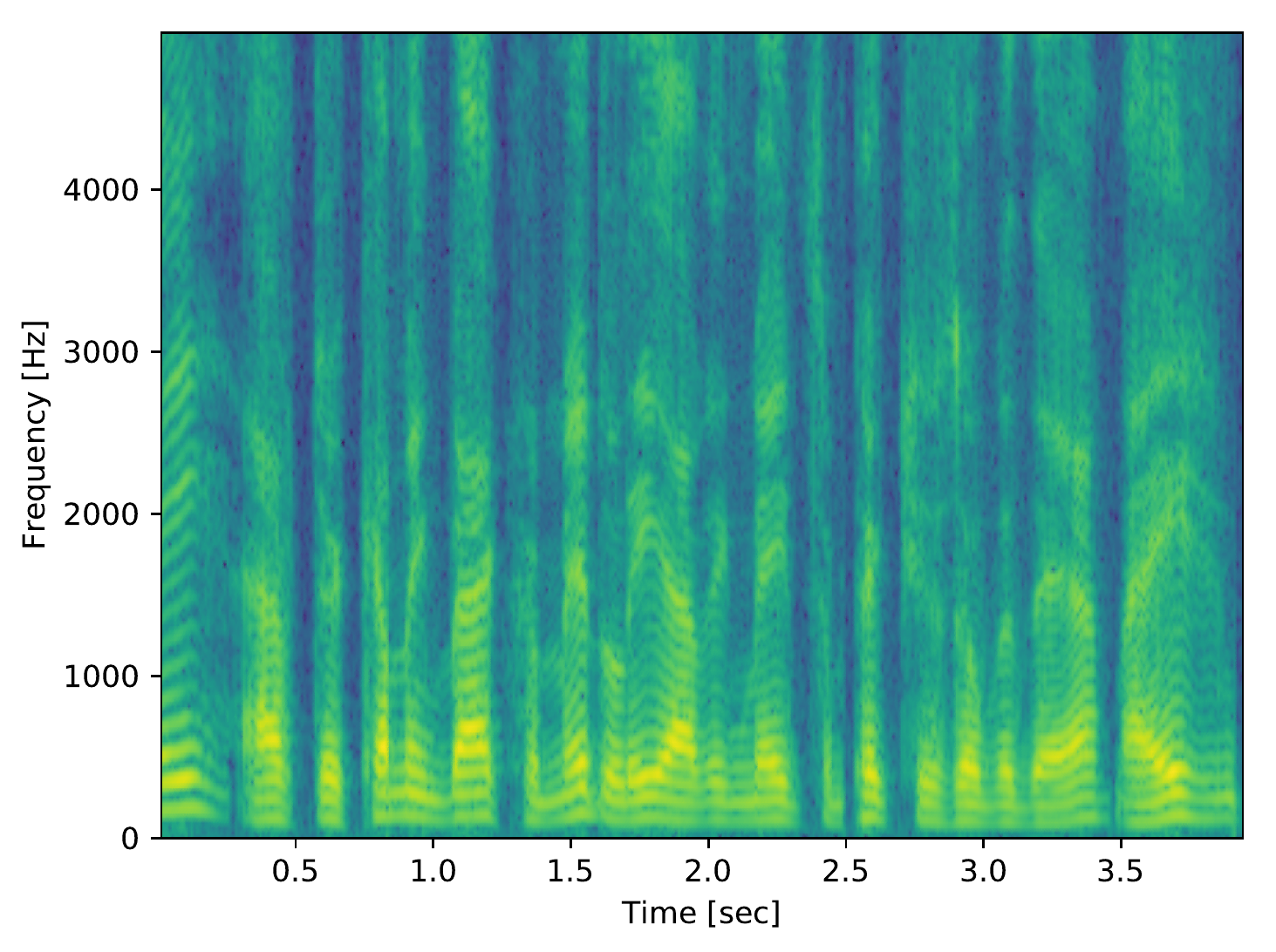} &
            \includegraphics[trim=0.0cm 0.0cm 0.0cm 0.0cm, clip=true, width=4.3cm]{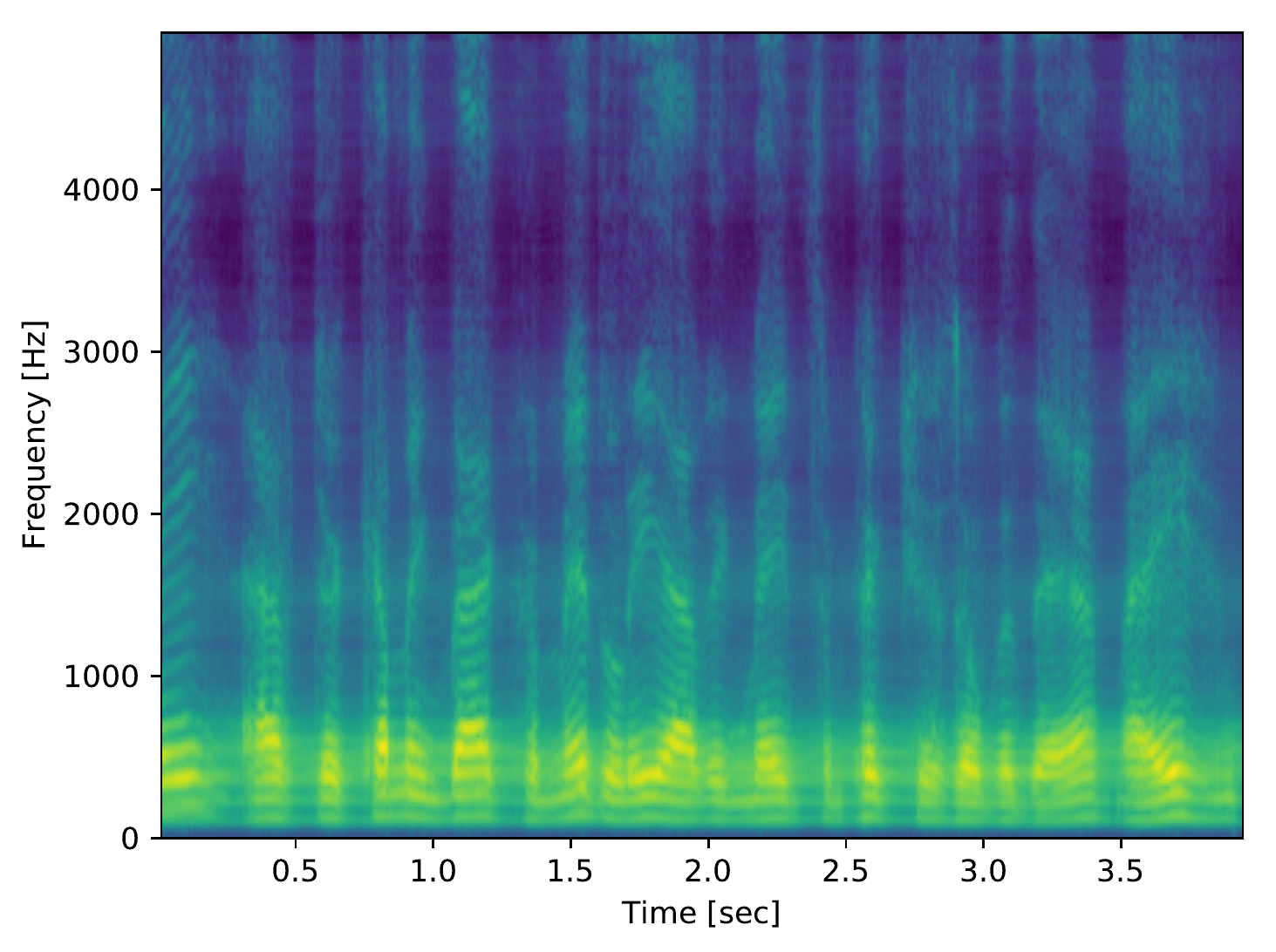}
            \\
            (a) Original signal before SE & (b) Original signal after SE
            \\
            \includegraphics[trim=0.0cm 0.0cm 0.0cm 0.0cm, clip=true, width=4.3cm]{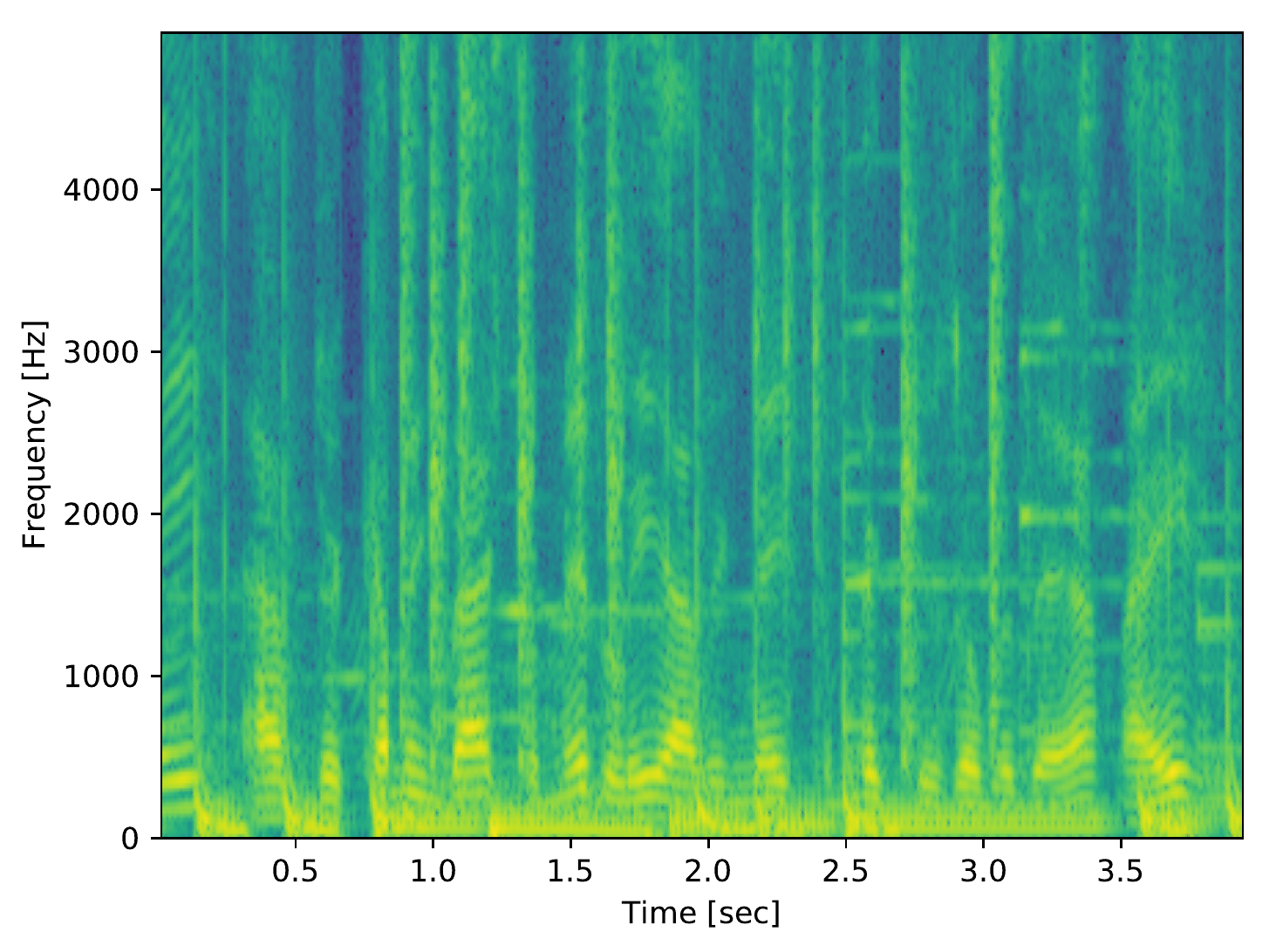} &
            \includegraphics[trim=0.0cm 0.0cm 0.0cm 0.0cm, clip=true, width=4.3cm]{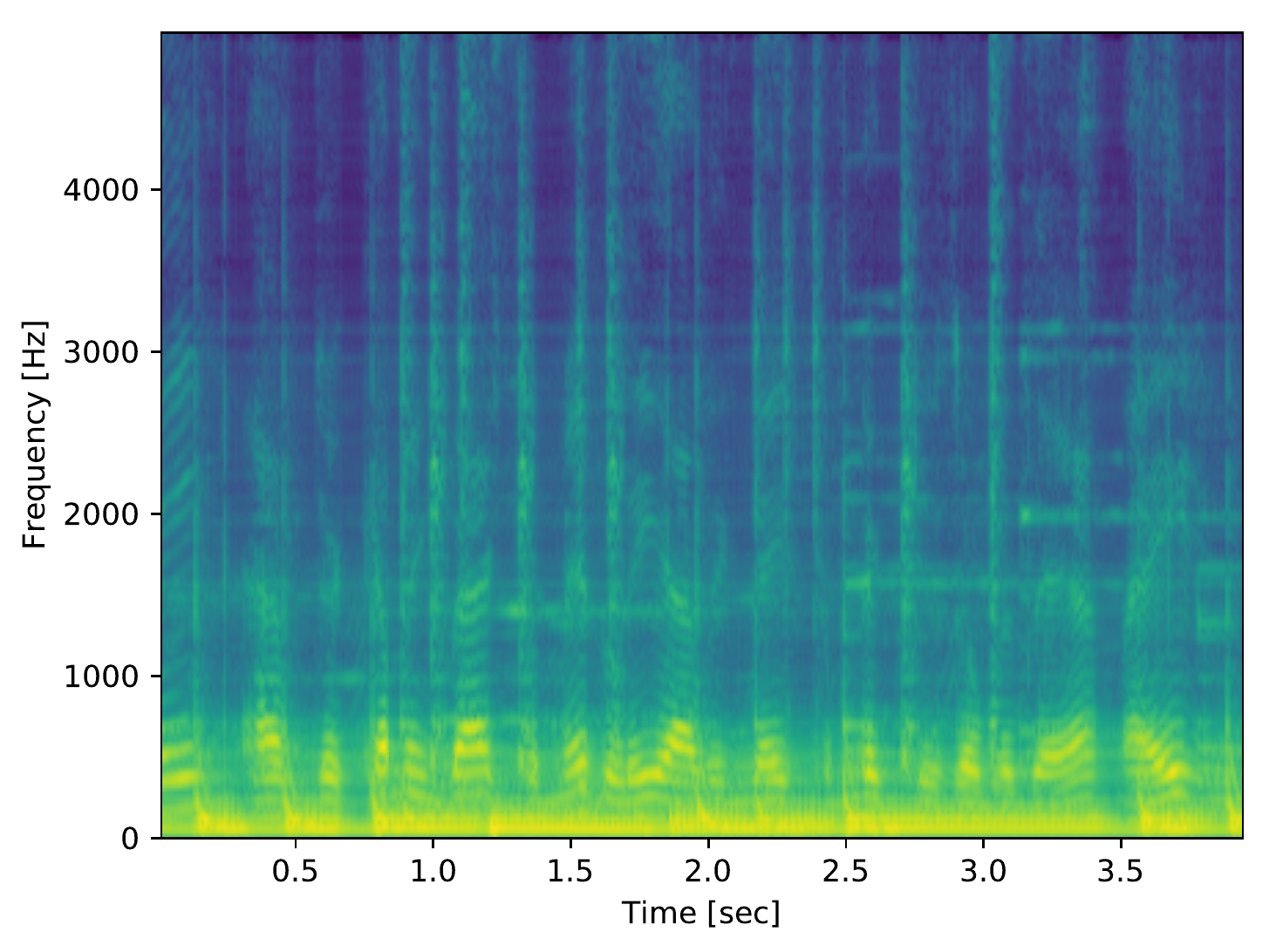} 
            \\ 
            (c) Noisy signal before SE & (d) Noisy signal after SE
        \end{tabular}
    \end{footnotesize}
\vspace{-0.0cm}
\caption{\textbf{Visualization of example spectrograms of the VoxCeleb dataset.}}
\vspace{-0.35cm}
\label{fig:STFT}
\end{figure}

\subsection{Impact of speech enhancement}

In this subsection, we investigate the impact of the masking-based speech enhancement (SE) for speaker verification (SV). 
Table \ref{ablation_SE} compares the results with and without the masking-based SE on the KID and VoxCeleb datasets, respectively. We used 2D-Res34-Fbank64 as a feature extractor with the softmax loss for training and the SAP for feature pooling. 
For both datasets, we observe that the FPM-based MSA improves the performance by using the SE model. This indicates that we can increase the robustness to acoustic distortions by jointly training the SV and SE models. 

In Fig. \ref{fig:STFT}, we visualize example spectrograms extracted from the VoxCeleb test set, where higher amplitudes are represented by brighter colors (yellow) while lower amplitudes are represented by darker colors (blue).
To do so, we used 2D-Res34-Spec160 as a feature extractor with the softmax loss for training and the SAP for feature pooling. 
Spec160 covers a frequency range of 0-5 kHz as explained in Section \ref{sec:implementation_details}, and a 4 s segment within an utterance is used.
Fig. \ref{fig:STFT}(a) shows the spectrogram extracted from a test file with the name ``id10300/8Sz2-IYJ2GA/00005.wav.'' 
The enhanced spectrogram of the sample is presented in (b). 
In (c), we visualize a spectrogram of the corrupted sample with music noise at SNR = 0 dB. Comparing (a) and (c), we can see that there are more horizontal and vertical lines in the noisy spectrogram (marked in yellow), which are generated by the music noise. 
Fig. \ref{fig:STFT}(d) shows the enhanced spectrogram of the corrupted sample. 
Compared to (c), the lines generated by the noise become darker (i.e., noise components become weaker) while the harmonic structures of speech remain bright.
This means that the SE model preserves the speech components while suppressing the noise components. We can see that, as expected, the SE model produces enhanced features even though it is jointly trained with the SV model without an explicit loss function.

\begin{table}[t]
\centering
\begin{footnotesize}
\caption{\textbf{Performance of three VAD models, averaged by five noise types, on the Aurora4 dataset.}}
\label{VAD_performance}
\vspace{-0.2cm}
\renewcommand{\arraystretch}{1.0}
\begin{tabular}{c|c||c|c|c|c}
\hline
Metric               & Model & -5 dB & 0 dB  & 5 dB & Average  \\ \hline
\multirow{3}{*}{AUC (\%)} & DNN   & 87.11 & 92.44 & 96.19 & 91.91 \\  
                     & LSTM  & 89.46 & 94.93 & 97.55 & 93.98 \\
                     & CLDNN & \textbf{91.01} & \textbf{95.70} & \textbf{97.82} & \textbf{94.84} \\ \hline
\multirow{3}{*}{EER (\%)} & DNN   & 19.85 & 14.29 & 9.36 & 14.50 \\  
                     & LSTM  & 18.24 & 12.19 & 8.14 & 12.86 \\  
                     & CLDNN & \textbf{16.42} & \textbf{11.02} & \textbf{7.55} & \textbf{11.66} \\ \hline
\end{tabular}
\vspace{-0.1cm}
\end{footnotesize}
\end{table}

\subsection{Impact of self-adaptive soft VAD}

In Table \ref{VAD_performance}, we present the results of three types of VAD models: DNN, LSTM, and CLDNN-based VADs (see Table \ref{VAD_architecture} for detailed architectures). 
For all models, Fbank64 features were used. 
We report the VAD performance in terms of the area under the ROC curve (AUC) and EER. 
For each SNR, we obtained the results for the five noise types, and averaged them as the final result.
The column ``Average'' denotes the overall average values over three SNRs.
We find that the CLDNN-based model performs best in all cases in terms of both metrics. 
The LSTM-based VAD is better than the DNN-based VAD, but worse than the CLDNN-based VAD. 
Note that all the models have similar number of parameters for a fair comparison.
This result is the same as that of \cite{Zazo2016}.

\begin{table}[t]
\centering
\begin{scriptsize}
\renewcommand{\arraystretch}{1.2}
\caption{\textbf{Impact of the selection of feature maps, where soft VAD is applied, on the KID dataset. The evaluation metric is the EER (\%).}}
\label{VAD_location}
\vspace{-0.1cm}
\setlength\tabcolsep{4pt}
\begin{tabular}{c|c|c|c||c|c|c|c}
\hline
No. & FPM & VAD & Feature map & S1-N6 & S2-N6 & S3-N6 & \textbf{S4-N6} \\ \hline
1 & $\times$            &  $\times$    & $\times$ & 18.83  &  12.04 & 8.85 & 8.68  \\ 
2 &   $\times$   & Hard (E) & $\times$ & 18.83  &  12.04 & 8.85 & 8.68  \\ 
3 & $\times$     & Hard (L) &  $\times$ & 17.85  &  11.04 & 8.71 & 8.60  \\ \hline
4  &  \checkmark  & $\times$ & $\times$ & 17.02  &  10.17 & 7.19 & 7.05  \\ 
5  &  \checkmark  & Hard (L) & $\times$ & 16.43  &  10.10 & 7.09 & 6.80  \\ \hline
6  &  \checkmark  & Soft (L) & $P_2$ &  19.24  &  12.08 & 8.69  & 8.84   \\ 
7  &  \checkmark  & Soft (L) & $\{P_k\}_{2\leq k\leq 3}$ & 19.08  &  11.91 & 8.77 & 8.64 \\ 
8  &  \checkmark   & Soft (L) & $\{P_k\}_{2\leq k\leq 4}$ &  16.53  &  10.33  & 7.50  & 7.33 \\ 
9 &  \checkmark  & Soft (L) & $\{P_k\}_{2\leq k\leq 5}$ &  16.29  &  10.07 & 7.40 & 7.15    \\ \hline
10 &  \checkmark  & Soft (D) & $\{P_k\}_{2\leq k\leq 5}$ &  17.21  &  10.97 & 8.05 & 7.71    \\ 
11 &  \checkmark  & Soft (C) & $\{P_k\}_{2\leq k\leq 5}$ &  15.89  &  9.67 & 7.21 & 7.04    \\ \hline
12 &  \checkmark  & SAS (D) & $\{P_k\}_{2\leq k\leq 5}$ &  15.98  &  9.93 & 7.61 & 7.09    \\ 
13 &  \checkmark  & SAS (L) & $\{P_k\}_{2\leq k\leq 5}$ &  \textbf{15.54}  &  \textbf{9.13} & \textbf{6.56} & 6.19    \\
14 &  \checkmark  & SAS (C) & $\{P_k\}_{2\leq k\leq 5}$ &  15.62  &  9.20 & 6.60 & \textbf{6.12}    \\ \hline
\end{tabular}
\vspace{-0.1cm}
\end{scriptsize}
\end{table}

When we combine FPM-based MSA and SAS-VAD, speech posteriors from VAD are multiplied to feature maps from FPM, as shown in Fig. \ref{fig:sync_network}.
As there are four feature maps extracted from the FPM (i.e., $P_2$, $P_3$, $P_4$, and $P_5$), we can select at most four feature maps where soft VAD is applied.
We performed experiments to investigate the impact of the selection of feature maps for the soft VAD. 
Table \ref{VAD_location} shows the results on the KID dataset. 
For SV, we use 2D-Res34-Fbank64 as a feature extractor with softmax loss and SAP. 
In the table, we report the results of four different test conditions: S1-N6, S2-N6, S3-N6, and S4-N6, which contain long non-speech regions (6 s) in noisy and reverberant environments. For enrollment, we used the test set of S4-N6.
The column ``Feature map'' indicates the selected feature maps for the soft VAD, where the mark ``$\times$'' indicates that all feature maps are not selected, that is, the soft VAD is not applied.
The column ``FPM'' and ``VAD'' denote the FPM-based MSA and the type of VAD, respectively.
In ``VAD,'' Hard stands for hard VAD which is a typical VAD making a hard decision based on a predefined threshold.
Soft and SAS indicate soft VAD and self-adaptive soft VAD (SAS-VAD), respectively.
E, D, L, and C in the parentheses stand for energy-, DNN-, LSTM-, and CLDNN-based VADs, respectively.

The first three rows show the results without the FPM-based MSA, where only $C_5$ is used. 
Hard (E) is not useful at all because the energy-based VAD is severely degraded by noise and reverberation.
On the other hand, the third system achieves lower EERs on all test conditions by applying the LSTM-based hard VAD. 
From the 4th to the 9th row, we can see the impact of the selection of feature maps for the LSTM-based soft VAD. Note that we used all feature maps, i.e., $\{P_2, P_3, P_4, P_5\}$, for the FPM-based MSA, and changed the feature maps where soft VAD is applied. 
The 4th row provides the result when the VAD is not applied. 
By comparing the 1st and 4th rows, we can observe that the FPM-based MSA itself deals with long non-speech intervals even without using VAD. 
From the 6th to the 9th row, we gradually increase the number of feature maps where the soft VAD is applied. 
When we apply the soft VAD to only $P_2$ or $\{P_2, P_3\}$, it degrades the SV performance. 
When we use all feature maps, i.e., $\{P_2, P_3, P_4, P_5\}$, the soft VAD improves the SV performance on S1-N6 and S2-N6, but slightly degrades the performance on S3-N6 and S4-N6.
These results indicate that the soft VAD performs best when it is applied to all feature maps.
We can also see that the effect of the soft VAD gradually increases as the duration of the speech segment decreases.
This suggests that the FPM-based MSA needs the soft VAD when the speech segment is too short (i.e., 1 s or 2 s), but it is somewhat robust to long non-speech segments when the speech segment is relatively long (i.e., 3 s or 4 s).

The 13th row shows the results when we apply the SAS-VAD to all feature maps using LSTM-based VAD.
This gives better results than the FPM-based MSA without VAD (in the 4th row) on all test conditions.
Comparing the 5th and 13th rows, we can see that the SAS-VAD outperforms the hard VAD.
From this result, we can conclude that our unified framework performs better than the conventional approach using separately pre-trained SV and VAD models.
Besides, we can achieve improved performance on all test conditions compared to the system using only the soft VAD (in the 9th row).
As the VAD network is adapted to the speaker verification data, it is sufficiently robust to long non-speech segments in noisy and reverberant environments. 
Therefore, we can increase the robustness of the SV system to long non-speech segments by using the SAS-VAD.

To compare the SV performance with different VAD models, we performed experiments for soft VAD and SAS-VAD, respectively.
In the case of the soft VAD (9th, 10th, and 11th rows), the SV performance is directly proportional to the VAD performance (see Table \ref{VAD_performance}). 
That is, the CLDNN-based model performs best and the LSTM-based model performs second best.
However, in the case of the SAS-VAD (12th, 13th, and 14th rows), both models show similar results. 
Even on S1-N6, S2-N6, and S3-N6, the LSTM-based model performs slightly better than the CLDNN-based model. 
Therefore, in the following experiments, we only consider the LSTM-based VAD.

\begin{table}[t]
\centering
\begin{scriptsize}
\renewcommand{\arraystretch}{1.2}
\caption{\textbf{Ablation results on the KID dataset (EER \%).}}
\label{combined_ablation}
\vspace{-0.1cm}
\setlength\tabcolsep{4pt}
\begin{tabular}{c|c|cccc||c|c|c|c}
\hline
No. & Feat                   & FPM & VAD & FL & SE & S1-N6 & S2-N6 & S3-N6 & \textbf{S4-N6} \\ \hline
1 & \multirow{6}{*}{FB} &  $\times$  &  $\times$  &  $\times$  &  $\times$  &  18.83   &  12.04     &  8.85     &  8.68       \\ 
                    2 &   & \checkmark    &  $\times$  &  $\times$  &  $\times$  &      17.02   &  10.17     &  7.19     &  7.05        \\ 
                    3 &   & \checkmark    &  $\times$  &  $\times$  &   \checkmark  &     15.34     &  8.97      &    6.66   &    6.52       \\ 
                    4 &  & \checkmark       & \checkmark   &   $\times$ & $\times$     &        15.54   &     9.13  &    6.56   &    6.19       \\ 
                    5 &   & \checkmark       &  \checkmark  &  \checkmark  &  $\times$  &    15.05      &  9.08     &  6.54   &    6.21       \\ 
                    6 &   &  \checkmark       &  \checkmark  &  \checkmark  &  \checkmark &    \textbf{14.52}     &  \textbf{7.67}     &  \textbf{5.22}     &  \textbf{5.19}       \\ \hline
7 & \multirow{5}{*}{Spec}  &   $\times$  &   $\times$  & $\times$   &  $\times$  & 20.09     &  11.93     &  8.69  &    8.24   \\ 
                     8 &  &  \checkmark   &   $\times$  &  $\times$  & $\times$   & 15.50     &    8.24   &    5.42   &    4.97   \\
                     9 &  & \checkmark     &  $\times$   &  $\times$  & \checkmark   & 14.72     &  7.67     &     5.21  &  4.76  \\ 
                     10 &  & \checkmark     &  \checkmark   &  \checkmark  & $\times$    & 15.09     &  8.09     &     5.25  &  4.70  \\
                     11 &  & \checkmark     &  \checkmark    &  \checkmark   & \checkmark  & \textbf{14.05}     &  \textbf{6.98}     &     \textbf{4.70}  &  \textbf{4.05}  \\ \hline
\end{tabular}
\end{scriptsize}
\end{table}

\begin{table*}[t]
\centering
\begin{footnotesize}
\renewcommand{\arraystretch}{1.1}
\caption{\textbf{Ablation results of the integrated model on the VoxCeleb dataset (EER \%).}}
\label{SAS-VAD-FPM}
\vspace{-0.2cm}
\begin{tabular}{ccc||ccccc|cccc}
\hline
\multirow{2}{*}{FPM} & \multirow{2}{*}{VAD+FL} & \multirow{2}{*}{SE}  & \multicolumn{5}{c|}{Fixed S \& Variable N}           & \multicolumn{4}{c}{Variable S \& Fixed N}        \\ \cline{4-12} 
                     &                      &                      & S4-N0   & \textbf{S4-N2}  & S4-N4  & S4-N6  & S4-N8  & S1-N6 & S2-N6 & S3-N6 & \multicolumn{1}{c}{\textbf{S4-N6}} \\ \hline
$\times$                    & $\times$                    & $\times$                      &     9.20    &  9.44      &    10.19    &  11.49      &    12.43    & 18.06      &     13.86  &    12.20   & 11.01                           \\ 
\checkmark                    & $\times$                    & $\times$                       &    8.38     &  8.75      & 9.52       &  10.79      &     11.87   &    16.10   &  12.34     &     10.69  &        9.27                    \\ 
\checkmark                    & \checkmark                    & $\times$                        &  7.91       &    7.42    &  7.77      &  8.67     &  9.76      &   15.97    &    12.04   & 10.07      &  8.73                          \\ 
\checkmark                     & $\times$                    & \checkmark                        &     7.65    &     7.35   &    7.71    &     8.34   &   8.97     & 15.94      &  12.05     &  10.44     &  8.68                           \\ 
\checkmark                     & \checkmark                     & \checkmark                        &    \textbf{7.33}     &     \textbf{7.11}   &    \textbf{7.52}    &   \textbf{8.02}     &   \textbf{8.38}      &  \textbf{14.36}    &    \textbf{10.76}   & \textbf{9.25}      & \textbf{8.25}                           \\ \hline
\end{tabular}
\end{footnotesize}
\end{table*}

\subsection{Impact of the integrated model}

In this section, we evaluate the performance of the integrated model consisting of all three models: 1) the SV model using the FPM-based MSA, 2) the VAD model using the advanced SAS-VAD, and 3) the SE model using the masking network.
As demonstrated above, the FPM-based MSA improves the robustness to short speech segments.
The VAD and SE models improve the robustness to long non-speech segments and acoustic distortions, respectively.
To achieve our goal, we combined all models in a unified framework and jointly trained them.
We conducted ablation studies to demonstrate the effectiveness of each model in the integrated model. 

Table \ref{combined_ablation} shows the results (EER \%) on the KID dataset using two acoustic features, Fbank64 and Spec160, which are abbreviated as FB and Spec, respectively. 
We used 2D-Res34 as a feature extractor with softmax loss and SAP.
The column ``VAD'' and ``FL'' stand for SAS-VAD and focal loss, respectively. 
The 1st row to the 6th row are the results with Fbank64.
Comparing the 4th and the 5th rows, we can see that the focal loss enhances the SAS-VAD, especially when the speech segment is short. 
We believe that this is because the focal loss helps address the speech/non-speech class imbalance in the SAS-VAD.
By reducing the relative loss for well-classified examples (i.e., with a frequent class), the focal loss focuses training on misclassified examples (i.e., with a rare class), thus dealing with class imbalance.
The 6th row gives the results of the final integrated model, which achieves the best performance among them. 
Compared to the baseline model in the 1st row, the integrated model shows a relative improvement of 22.89\% in EER on S1-N6.
We can conclude that all components help each other to improve SV performance on the challenging scenario, where the input utterance contains short speech segments and long non-speech segments degraded by noise and reverberation.
The last five rows are the results with Spec160. 
We can observe the same trend as in the results with Fbank64.
Compared to the case of Fbank64, the baseline model achieves similar performance on S2-N6, S3-N6, and S4-N6, but it shows worse performance on S1-N6.
However, when the FPM-based MSA is applied, the model using Spec160 shows a larger improvement than the model using Fbank64, thereby achieving higher performance on all test conditions.
The last row shows the results of the final integrated model using Spec160, which are the best among all models.
Fig. \ref{fig:det} shows detection error trade-off (DET) plots of the five models: the baseline (in the 7th row), the model with FPM-based MSA (in the 8th row), the model with SE (in the 9th row), the model with SAS-VAD (in the 10th row), and the final integrated model (in the 11th row).
The figure shows the same trend as Table \ref{combined_ablation}.

\Figure[!t]
(topskip=0pt, botskip=0pt, midskip=0pt)[trim=4.65cm 9.5cm 5.15cm 10.2cm, clip=true, width=0.46\textwidth]{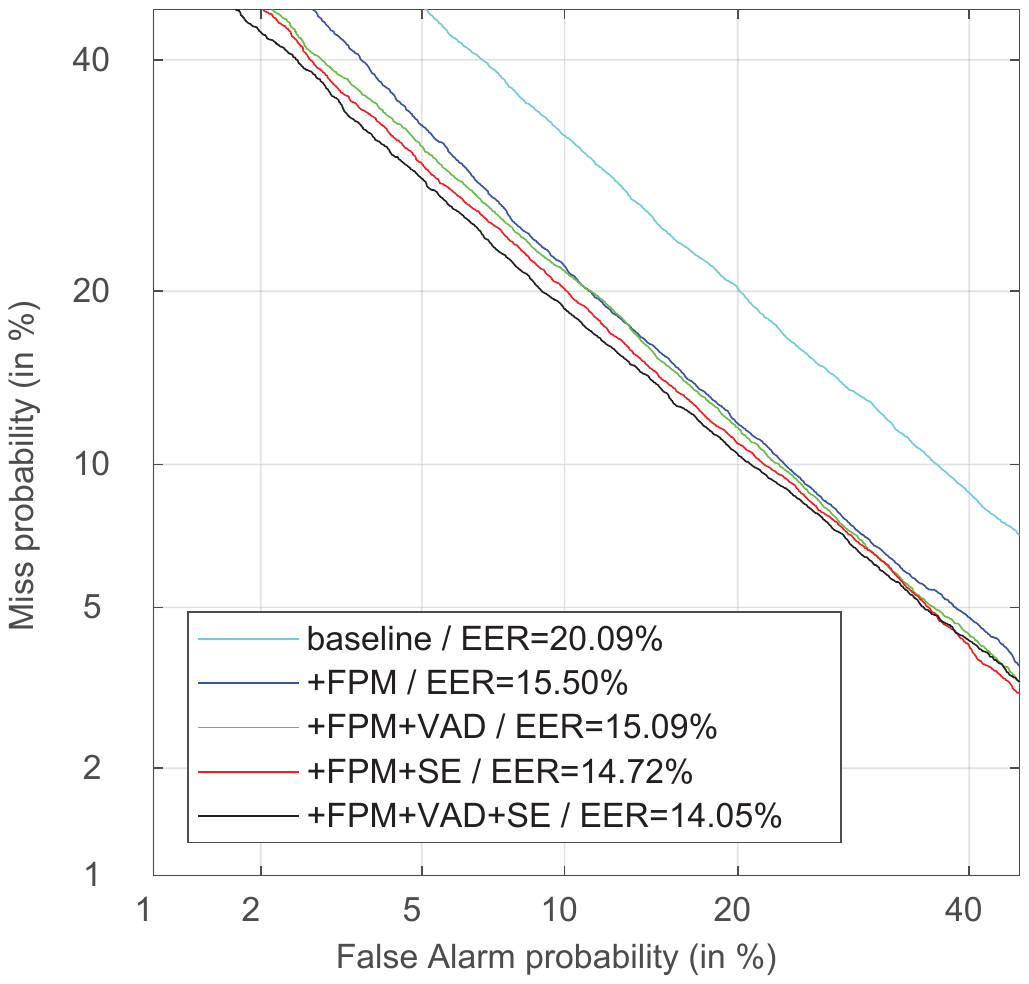}
   {\textbf{The DET curves of the proposed systems with FPM-based MSA, VAD, and SE, on S1-N6 of the KID dataset.}\label{fig:det}}

\begin{table*}[t]
\centering
\begin{footnotesize}
\renewcommand{\arraystretch}{1.2}
\caption{\textbf{EER (\%) comparison with the \textit{i}-vector/PLDA and state-of-the-art systems on the KID dataset.}}
\vspace{-0.2cm}
\label{final_table}
\begin{tabular}{c|c||ccccc|c|cccc|c}
\hline
\multirow{2}{*}{System} & \multirow{2}{*}{Feat} & \multicolumn{6}{c|}{Fixed S \& Variable N} & \multicolumn{5}{c}{Variable S \& Fixed N}  \\ \cline{3-13} 
                        & & S4-N0  & \textbf{S4-N2}  & S4-N4  & S4-N6  & S4-N8 & Avg. & S1-N6     & S2-N6    & S3-N6    & \textbf{S4-N6} & Avg.    \\ \hline
\textit{i}-vector/PLDA  & MFCC60         & 10.41  & 11.75  & 12.27  & 13.64  & 14.95 & 12.60 & 28.21     & 21.24    & 16.15    & 14.20  & 19.95   \\ 
TDNN (SM+SP) \cite{Snyder2017} & Fbank64                      &   8.23     &    8.58    &   8.90     &  9.58      &     10.40  & 9.14  &       22.61    &      14.73    &      10.98    &  10.05 & 14.59        \\ 
TDNN (SM+ASP) \cite{Okabe2018} & Fbank64                       &  8.10      &  8.35      &  8.78      &     9.69   & 10.49 & 9.08 &   22.82       &  14.63        &  10.82        &   9.71 & 14.50       \\ 
2D-Res34 (ASM+SPE) \cite{Jung2019} & Fbank64                       &   6.95     &  7.03      &  7.49      & 7.81       &    8.36  & 7.53  &       23.98    &      11.18    &      7.89    &   6.99 & 12.51       \\ 
Proposed (SM+SAP) & Spec160                       &   4.17     &    3.95    &  4.03      &  4.59      &  4.88   & 4.32    &    14.05       &   6.98       &  4.70  &  4.05 & 7.44        \\
\textbf{Proposed (ASM+SAP)} & Spec160                       &  \textbf{3.98}      &  \textbf{3.73}      &     \textbf{3.89}   &    \textbf{4.15}    &  \textbf{4.41} & \textbf{4.03}     &      \textbf{12.94}     & \textbf{5.85}         &  \textbf{4.08}        &   \textbf{3.90}   & \textbf{6.69}    \\ \hline
\end{tabular}
\end{footnotesize}
\end{table*}

Table \ref{SAS-VAD-FPM} lists the EERs of five models using Spec160 on the VoxCeleb dataset. 
All models used the attentive statistics pooling (ASP) \cite{Okabe2018} layer for feature pooling, which uses not only the weighted mean but also the weighted standard deviation of feature vectors. 
The column ``VAD+FL'' stands for SAS-VAD with focal loss.
Different from Table \ref{combined_ablation}, we performed ablation studies of the integrated model using more test conditions.
In the left five conditions, denoted as ``Fixed S \& Variable N,'' we fixed the length of speech segments (S) as 4 s and changed the length of non-speech segments (N). The test set of S4-N2 was used for enrollment.
As in the case of the KID dataset, each model improves SV performance and the integrated model using all models performs best.
We can see that the results on S4-N2 are better than those on S4-N0 when VAD or SE is applied, which is not consistent with our intuition.
We speculate that this is because the length difference between enrollment and verification utterances affects SV performance to some extent, and both utterances have the same length in S4-N2 (i.e., the total length of 6 s).
In the right four conditions, denoted as ``Variable S \& Fixed N,'' we fixed the length of non-speech segments (N) as 6 s and changed the length of speech segments (S).
The test set of S4-N6 was used for enrollment.
It can be observed that the final integrated model achieves the best performance on all test conditions, which is the same as in the KID dataset.

In Table \ref{final_table}, we compare the proposed system with the \textit{i}-vector/PLDA system and state-of-the-art deep speaker embedding systems including \textit{d}-vector and \textit{x}-vector-based systems. All results were obtained by our own implementation. 
The table shows the average EERs for both ``Fixed S \& Variable N'' and ``Variable S \& Fixed N,'' respectively.
In the column ``System,'' the first and second terms in the parentheses indicate the loss function and global pooling layer, respectively. Here, SM and ASM stand for softmax and angular softmax (A-Softmax) loss functions, respectively. 
Specifically, for ASM, we used the combination of A-Softmax loss and ring loss with the same settings as in \cite{Jung2019}. 
SP and SPE stand for statistics pooling and spatial pyramid encoding, respectively. 
In the case of \textit{i}-vector/PLDA and TDNN (i.e., \textit{x}-vector-based system), we used the same architectures as in Table \ref{TDNN_ResNet_comparison}. 
Different from the original \textit{x}-vector-based system, we did not use data augmentation for a fair comparison.
In the case of the 4th system (i.e., \textit{d}-vector-based system), we followed the same approach as in our previous work \cite{Jung2019}.
The proposed system is the system of 11th row in Table \ref{combined_ablation}, which includes all components (FPM, VAD+FL, and SE).
We obtained the best results by applying the ASM to the proposed system, which is given in the last row.
Our best system outperforms the \textit{i}-vector/PLDA system with a relative improvement of 68.02\% and 66.47\% in terms of the average EER for ``Fixed S \& Variable N'' and ``Variable S \& Fixed N,'' respectively.

In the case of the TDNN-based system, the ASP layer performs better than the SP layer on S4-N0, S4-N2, and S4-N4.
However, when the length of non-speech segments is too long (i.e., S4-N6 and S4-N8), the ASP layer degrades the SV performance. According to the results in ``Variable S \& Fixed N,'' we can observe that the ASP layer degrades the performance of TDNN-based system when the speech segment is very short (i.e., S1-N6).
From these results, we can conclude that the attention-based pooling does not work well when the speech segments are too short or non-speech segments are too long. 
Compared to the TDNN-based system using the ASP layer, our best system shows a relative improvement of 55.62\% and 53.86\% for ``Fixed S \& Variable N'' and ``Variable S \& Fixed N,'' respectively.
Even though the attention-based pooling is used in the proposed system, there is no performance degradation when the speech segments are too short or non-speech segments are too long.
We argue that this is because FPM-based MSA, SAS-VAD, and masking-based SE improve the robustness of the proposed system to short speech segments and long non-speech segments in noisy and reverberant environments.

\section{Conclusion}
\label{sec:conclusion}
In this study, we set two goals for speaker verification (SV) : an SV model should be robust to short speech segments and long non-speech segments, especially in noisy and reverberant environments.
The FPM-based MSA was applied to the SV model to deal with short speech segments, and
the SAS-VAD algorithm was used to deal with long non-speech segments.
For the SAS-VAD, the focal loss was adopted to address the class imbalance problem, and the 1D-CNN-based synchronizer was proposed to combine the SAS-VAD and the FPM-based MSA.
The masking-based speech enhancement (SE) was applied to further increase the robustness to acoustic distortions, especially noise and reverberation.
To achieve the goals simultaneously, we proposed a novel unified deep learning framework that integrates SV, VAD, and SE models into a single model and jointly trains the integrated model. 
Extensive experiments were conducted on two datasets: Korean indoor (KID) and VoxCeleb datasets, which are corrupted by noise and reverberation. 
The effectiveness of the MSA was demonstrated using three types of feature extractors and two types of acoustic features.
Also, several ablation studies were conducted to investigate the impact of each component in the integrated model.
The proposed system obtained the best results on various test conditions, including those with short speech segments and long non-speech segments, degraded by noise and reverberation.
Especially, it outperformed the conventional \textit{i}-vector/PLDA system with a relative improvement of approximately 67\% on the KID dataset. 
By jointly training the entire network in an end-to-end manner, we obtained better results than the conventional approach using separately pre-trained models.
We also provided a detailed overview of deep speaker embedding learning, in addition to the experiments.
In the future, we plan to develop a sophisticated approach to automatically find the optimal speech threshold in the SAS-VAD, instead of just using a fixed value. 
Also, we will figure out how to improve the computational efficiency of the SAS-VAD, since it does not discard non-speech frames.
Besides, in this paper, we mainly focused on how to deal with short speech segments and long non-speech segments. Thus, we plan to extend our work by using more advanced SE models and compare with other state-of-the-art SE approaches.

\begin{IEEEbiography}[{\includegraphics[width=1in,height=1.25in,clip,keepaspectratio]{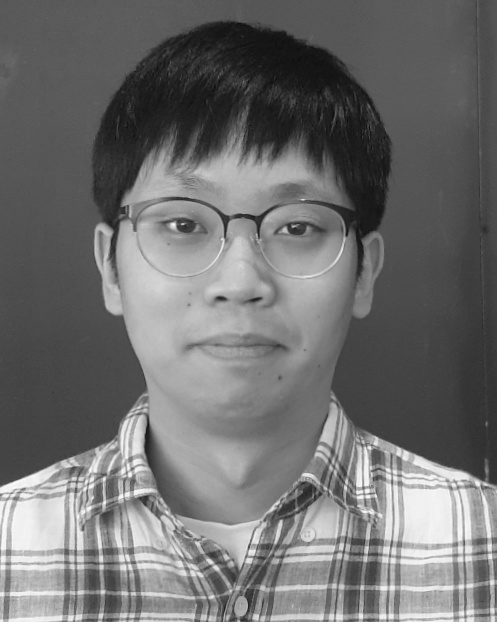}}]{YOUNGMOON JUNG} received the B.S. degree in electronic engineering from Sogang University, Seoul, South Korea, in 2016. He is currently pursuing the Ph.D. degree in electrical engineering with Korea Advanced Institute of Science and Technology
(KAIST), Daejeon, South Korea. 
His research interests include deep learning and signal processing for speaker recognition, speech synthesis, speech enhancement, and voice activity detection.

\end{IEEEbiography}

\begin{IEEEbiography}[{\includegraphics[width=1in,height=1.25in,clip,keepaspectratio]{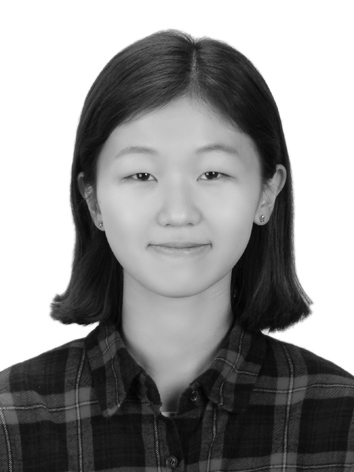}}]{YEUNJU CHOI} received the B.S. and M.S. degrees in electrical engineering from Korea Advanced Institute of Science and Technology (KAIST), Daejeon, South Korea, in 2016 and 2018, respectively, where she is currently pursuing the Ph.D. degree. Her research interests include deep learning and signal processing for speech synthesis, voice conversion, speaker verification, spoofing detection, and speech quality prediction.

\end{IEEEbiography}

\begin{IEEEbiography}[{\includegraphics[width=1in,height=1.25in,clip,keepaspectratio]{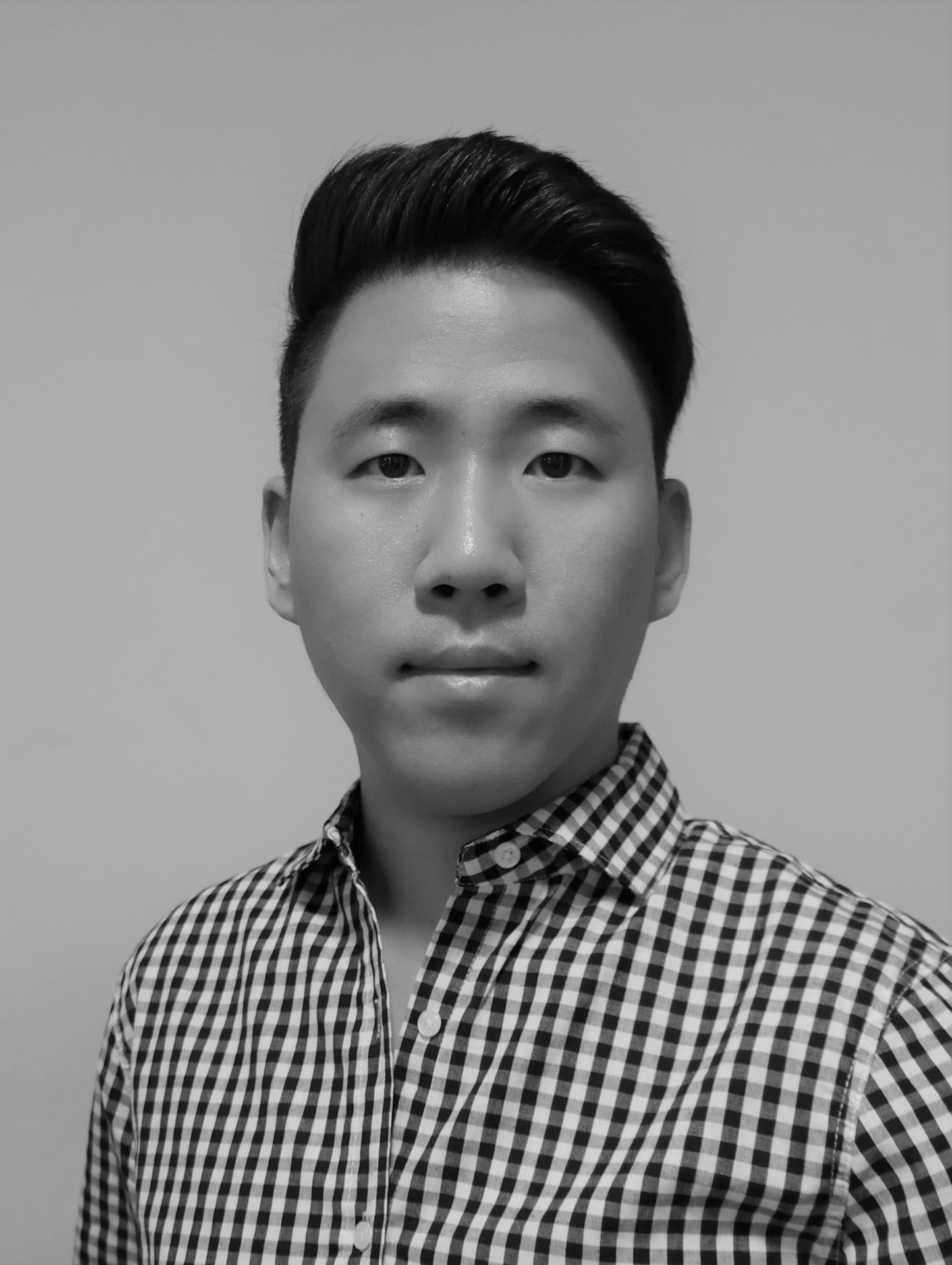}}]{HYUNGJUN LIM} received the B.S. degree in electronic engineering from Chung-Ang University, Seoul, South Korea, in 2013. He is currently pursuing the Ph.D. degree in electrical engineering with Korea Advanced Institute of Science and Technology
(KAIST), Daejeon, South Korea. 
His research interests include machine learning and signal processing for wake-up word detection, speaker verification, and sound event classification.

\end{IEEEbiography}

\begin{IEEEbiography}[{\includegraphics[width=1in,height=1.25in,clip,keepaspectratio]{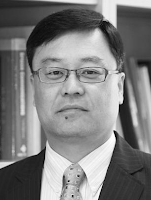}}]{Hoirin Kim} (M'92) received the B.S. degree in electronics engineering from Hanyang University, Seoul, South Korea, in 1984, and the M.S. and Ph.D. degrees in electrical engineering from Korea Advanced Institute of Science and Technology (KAIST), Daejeon, South Korea, in 1987 and 1992, respectively.

From 1987 to 1999, he was a Senior Researcher with the Spoken Language Processing Laboratory, Electronics and Telecommunications Research Institute, South Korea. From 1994 to 1995, he was on leave with ATR-ITL, Kyoto, Japan, as a Visiting Researcher. From 2000 to 2009, he was an Associative Professor with the Information and Communications University, South Korea. From 2006 to 2007, he was on leave with the INC, University of California at San Diego, San Diego, USA, as a Visiting Scholar. Since 2009, he has been a Professor with the School of Electrical Engineering, KAIST.
His research interests include signal processing for speech and speaker
recognition, speech synthesis, audio indexing and retrieval, and spoken language
processing.
\end{IEEEbiography}

\EOD

\end{document}